\documentclass[10pt]{article}
\usepackage{latexsym}
\usepackage{sparticles}
\usepackage{feynmf}
\def \BE {\begin{equation}}
\def \EE {\end{equation}}
\def \BEA {\begin{eqnarray}}
\def \EEA {\end{eqnarray}}
\def \CR {\nonumber \\}
\def \zer {^{(0)}}
\def \one {^{(1)}}
\def \two {^{(2)}}

\def \e {{\epsilon}}

\begin{document}
\title{Joint statistics of amplitudes and phases in Wave Turbulence}
\author{Yeontaek Choi$^*$, Yuri V. Lvov$^\dagger$ and Sergey Nazarenko$^*$
\\
$^*$
Mathematics Institute, The University of Warwick,  Coventry, CV4 7AL, UK \\
$^\dagger$ Department of Mathematical Sciences, Rensselaer
Polytechnic Institute, Troy, NY 12180 }

\maketitle

\begin{abstract}
Random Phase Approximation (RPA) provides a very convenient 
 tool  to study the ensembles of weakly interacting
waves, commonly called Wave Turbulence.
In its traditional
formulation, RPA assumes that phases of interacting waves are random
quantities but it usually ignores randomness of their amplitudes.
Recently, RPA was generalised in a way that takes into account the
amplitude randomness and it was applied to study of the higher momenta
and probability densities of wave amplitudes.  However, to have a
meaningful description of wave turbulence the RPA properties assumed
for the initial fields must be proven to survive over the nonlinear
evolution time, and such a proof is the main goal of the present
paper. We derive an evolution equation for the full 
probability density function which contains the complete information about 
the joint statistics of all wave amplitudes and phases.  We show that,
for any initial statistics of the amplitudes, the phase factors remain
statistically independent uniformly distributed variables.
If in addition the initial amplitudes are also independent variables
(but with arbitrary distributions) they will remain independent when
considered in small sets which are much less than the total number of modes.
However, if the size of a set is of order of the total number of modes
then the joint probability density for this set is not factorisable into
the product of one-mode probabilities. In the other words, the modes
in such a set are involved in a ``collective'' (correlated) motion.  
We also study new type of correlators describing
the phase statistics.
\end{abstract}

\section{Introduction}

Wave Turbulence (WT) is a common name for the fields of dispersive
waves which are engaged in stochastic weakly nonlinear interactions
over a wide range of scales. Plentiful examples of WT are found in
oceans, atmospheres, plasmas and Bose-Einstein
condensates~\cite{ZLF,Ben,GS,Newell68,Zakfil,hasselman,DNPZ}.
Roughly, there have been three major approaches to derive the WT
theory, one based on a diagrammatic approach
\cite{wyld,lvovzakh,llnz}, the second based on cumulant expansions
\cite{Ben, Newell68,davidson, DNPZ} and the third one, the random
phase approximation (RPA) \cite{ZLF,GS,Zakfil}.

The diagrammatic approach was developed in a field theoretical spirit
based on the Wyld's technique \cite{wyld}.  This method introduces an
artificial Gaussian forcing for which a zero limit is taken at the end
of the derivation. It is usually said that the statistical properties
of this force (Gaussianity) do not affect the statistical properties
of the resulting WT state which will be determined by the nonlinear
properties only. However, such independence of the WT state on the
statistics of the ``seed'' forcing is not obvious
because the limit of small nonlinearity is taken
before the limit of small force, i.e. the force  remains 
much greater than the nonlinearity.  In particular, when the nonlinearity
parameter is strictly zero, the Wyld technique gives a Gaussian steady state
which is clearly an artefact of this method because for linear systems
statistics of the wave amplitudes remain the same as in the initial
condition and, therefore, can be arbitrary. The question if any nonlinearity,
no matter how small, can break this dependence of the steady state on the
initial conditions still has not been answered in the literature.
 Thus, the diagrammatic
approach, although a very efficient way to build the perturbation
expansion, needs to be expanded to include non-Gaussian ``seed'' force
in order to see to what extent the results are not sensitive to the
force statistics. However, some elements of the Wyld technique will be
used in the present paper, not as a complete description but rather as
an auxiliary aid in writing out complicated terms.

The cumulant expansion approach  differs from the other methods by
working directly with the continuous Fourier transforms corresponding
to the infinite coordinate space without introducing a finite box as
an intermediate step. The main idea here is that, although the Fourier
transform is ill-defined for the wave fields corresponding to
homogeneous turbulence, it is well defined for the cumulants provided
that the correlations decay rapidly enough in the coordinate space.
The cumulant method is very elegant for describing the spectra and the
multiple-point moments the points in which are not ``fused'' (i.e. all
different).  However, some important statistical quantities involve
fused moments and they are hard (if at all possible) to define without
introducing a finite box as an intermediate step.  For example, one of
such objects, $\langle |a|^4 \rangle$, is important because it
describes intensity of fluctuations of the $k$-space distribution of
energy $|a|^2 $, namely $\delta E = \sqrt{\langle |a|^4 \rangle -
\langle |a|^2 \rangle^2} $ (see \cite{ln}).  Furthermore, non-decaying in the
$x$-space correlations tend to naturally develop over the nonlinear
time \cite{ln} and it is not clear what wave fields these correlations
correspond to within the cumulant approach.

RPA approach has been by far most popular technique due to its clear
intuitive content. However, this approach has occasionally been
downgraded to just a convenient way of interpreting the results of a
more rigorous technique based on the cumulant expansions.  It happened 
because RPA,  being widely used by
physicists,  had not been formulated rigorously. In particular, it
is typically assumed that the phases evolve much faster than
amplitudes in the system of nonlinear dispersive waves and, therefore,
the averaging may be made over the phases only ``forgetting'' that the
amplitudes are statistical quantities too (see
e.g. \cite{ZLF}).  This statement become less obvious if one takes
into account that we are talking not about the linear phases $\omega
t$ but about the phases of the Fourier modes in the interaction
representation.  Thus, it has to be the nonlinear frequency
correction that helps randomising the phases \cite{zs}.
On the other hand, for three-wave systems (considered in this paper)
the period associated with the nonlinear frequency correction
is of the same $\epsilon^2$ order in small nonlinearity $\epsilon$
as the nonlinear  evolution time and, therefore, phase randomisation
cannot occur faster that the nonlinear evolution of the amplitudes.
One could hope that the situation is better for 4-wave systems
(not considered here) because the nonlinear frequency correction
is still $\sim \epsilon^2$ but the nonlinear evolution appears
only in the  $\epsilon^4$ order. However, in order to make the
asymptotic analysis consistent, such  $\epsilon^2$ correction
has to be removed from the interaction-representation amplitudes
and the remaining phase and amplitude evolutions are, again,
at the same time scale (now $1/\epsilon^4$).
This picture is confirmed by the numerical simulations of 
the 4-wave systems \cite{zakhpush,cln} which indicate that the
nonlinear phase evolves at the same timescale as the amplitude.
Thus, to proceed theoretically one has to start with phases which
are already random (or almost random) and hope that this randomness
is preserved over the nonlinear evolution time. In most of the 
previous literature such preservation was assumed but not proven.
The goal of this paper will be to study the extent to which
such an assumption is valid. 

Another goal of this paper is to make RPA 
formulation more consistent by taking into account that both phases 
and the amplitudes are random variables. Indeed, even if one starts 
with a wavefield which has random phases but deterministic
amplitudes, as it is typically done in numerical simulations,
 the amplitudes will get randomised because the nonlinear term
producing their evolution contains (random) phase factors.
Preliminary steps were 
 recently done in \cite{ln,cln} where we assumed that all the phases and
the amplitudes in the {\em initial} wavefield are random variables
independent of each other and that the phase factors are uniformly
distributed on the unit circle on the complex plane.  We kept the
same acronym RPA but re-interpreted it as ``Random Phases and
Amplitudes'', reflecting the fact that, first, the amplitudes are also
random and, second, that it is not an ``approximation'' but rather an
assumed property of the {\em initial} field. Such a generalised RPA
was used in \cite{ln} to study the evolution of the higher moments of
the Fourier amplitudes and in \cite{cln} to study their
``one-mode'' PDF.  In fact, this form of RPA is more general than
the cumulant approach because it can handle fields with long
correlation lengths which appear to be important for intermittency
\cite{cln}.

Of course, for such an analysis to be trustworthy one should prove that
the RPA properties hold over the nonlinear time and not just for
the initial fields.  Such mathematical validation of the RPA method 
will be in the focus of the present paper. To do this we will have
to study the full joint PDF which involves the complete statistical
information about the system, including the multi-mode
correlations of both the amplitudes and the phase factors. 
We will derive an evolution equation for such PDF and we will show that it
is identical to the equation obtained for the excitations in anharmonic
crystals originally obtained by Peierls \cite{peierls} and later
reproduced by Brout and Prigogine \cite{bp} and Zaslavski and Sagdeev
\cite{zs}. All these works were restricted to considering a quite
narrow class of interaction Hamiltonians arising from a 
potential energy, i.e. depending on the coordinates but not momenta.
These class does not include a large number of interesting WT systems,
e.g. the capillary, internal, Rossby and Alfven waves. It is remarkable,
therefore, that the Peierls equation turns out to be universal for the general
class of three-wave systems, as it is shown in the present paper.
Further, we use this equation to validate an ``essential'' RPA formulation, i.e.
approximate RPA which holds only up to a certain order in nonlinearity
and discreteness, but which is sufficient for the WT closure.
This
validation gives RPA technique a status of a rigorous approach which,
due to the simplicity of its premises, is a winning tool for the
future theory of non-Gaussianity of WT, its intermittency and
interactions with coherent structures.

In addition to the mathematical validation of RPA, we will also
develop WT further by considering new statistically important
quantities. For a long time, describing and predicting the energy
spectra was the only concern in WT theory. Recently, we presented a
description of the higher order statistics of the one-point Fourier
correlators in terms of their moments and PDF's. They describe the
k-space ``noise'', i.e. the fluctuations of the mode energy about its
mean value given by the energy spectrum. We also showed PDF's have a
long algebraic tail which indicates presence of intermittency in WT
fields. The present paper deals with phases, and we will therefore
introduce and study some new correlators which will allow to describe
the phase statistics directly. Such a description will compliment the
mathematical validation of the RPA because it yields to a physical
answer on how initially correlated phases can get de-correlated in the
first place.

\section{Fields with Random Phases and Amplitudes. }
Let us consider a wavefield $a({\bf x}, t)$ in a periodic cube of with
side $L$ and let the Fourier transform of this field be $a_l(t)$ where
index $l {\in } {\cal Z}^d$ marks the mode with wavenumber $k_l = 2
\pi l /L$ on the grid in the $d$-dimensional Fourier space.  For
simplicity let us assume that there is a maximum wavenumber $k_{max}$
(fixed e.g. by dissipation) so that no modes with wavenumbers greater
than this maximum value can be excited.  In this case, the total
number of modes is $N = (k_{max} / \pi L)^d$. Correspondingly, index
$l$ will only take values in a finite box, $l \in {\cal B}_N \subset
{\cal Z}^d$ which is centred at 0 and all sides of which are equal to
$k_{max} / \pi L = N^{1/3}$.  To consider homogeneous turbulence, the
large box limit $N \to \infty $ will have to be taken.
\footnote{
It is easily to extend
the analysis to the infinite Fourier space,
$k_{max} = \infty$.
In this case, the full joint PDF would still have
to be defined as a $N \to \infty$ limit of an
$N$-mode PDF, but this limit would have to be taken
in such a way that both  $k_{max}$ and the density of the Fourier modes
tend to infinity simultaneously.}

Let us write the complex
$a_l$ as $a_l =A_l \psi_l $ where $A_l$ is a real positive
amplitude and $\psi_l $ is a phase factor which takes values on
${\cal S}^{1} $,
a unit circle centred at zero in the complex plane. Let us
define the $N$-mode joint PDF ${\cal P}^{(N)}$ as the probability for the
wave intensities $A_l^2 $ to be in the range $(s_l, s_l +d s_l)$
and for  the phase factors $\psi_l$ to be on the unit-circle
segment between $\xi_l$
 and   $\xi_l + d\xi_l$ for all $l \in {\cal B}_N$.
 In terms of this PDF, taking the averages
will involve integration over all the real positive $s_l$'s and
along all the complex unit circles of all $\xi_l$'s,

\BEA \langle f\{A^2, \psi \} \rangle
= \left(
 \prod_{ l {\cal 2 B}_N }  \int_{{\cal R}^{+} } ds_l \oint_{{\cal S}^{1} }
|d \xi_l| \right) \;  {\cal P}^{(N)} \{s, \xi \}
f\{s, \xi \} \label{pdfn} \EEA
where notation $f\{A^2,\psi\}$ means that $f$ depends on all
$A_l^2$'s and  all $\psi_l $'s in  the  set $\{A_l^2,
\psi_l; l {\cal 2 B}_N \}$ (similarly, $\{s, \xi \}$ means $\{s_l,
\psi_l; l \in  {\cal B}_N \}$, etc). The full PDF that contains the
complete statistical information about the wavefield  $a({\bf x}, t)$
in the infinite $x$-space can be understood 
as a large-box limit
$${\cal P} \{ s_k, \xi_k \}  =  \lim_{N \to \infty}
{\cal P}^{(N)} \{s, \xi \},
$$
i.e. it is a functional acting on the continuous functions
of the wavenumber, $s_k$ and $\xi_k$.
In the the large box
limit there is  a path-integral version of (\ref{pdfn}),
\BE \langle f\{A^2, \psi \} \rangle =
 \int {\cal D}s \oint
|{\cal D} \xi| \;  {\cal P} \{s, \xi \}  f\{s, \xi \}
\label{mean-path} \EE
The full PDF defined above involves all $N$ modes (for either
finite $N$ or in the $N \to \infty$ limit). By integrating out
all the arguments except for chosen few, one can have
reduced statistical distributions. For example, by 
integrating over all the angles and over all but $M$ amplitudes,we have
an ``$M$-mode'' amplitude PDF,
\BE
{\cal P}_{j_1, j_2, \dots , j_M} = \left(
\prod_{ l \ne j_1, j_2, \dots , j_M }  \int_{{\cal R}^{+} } ds_l 
\prod_{ m {\cal 2 B}_N } 
\oint_{{\cal S}^{1} }
|d \xi_m| \; \right) {\cal P}^{(N)} \{s, \xi \},
\EE
which depends only on the $M$ amplitudes marked by labels
$j_1, j_2, \dots , j_M  {\cal 2 B}_N$.

\subsection{Definition of an ideal RPA field}

Following  the approach of  \cite{ln,cln}, we now define a
``Random Phase and Amplitude'' (RPA) field.  
We say that the field $a$ is of RPA type if it
possesses the following statistical properties:

\begin{enumerate}
\item All amplitudes $A_l$  and their phase factors $\psi_l $  are
independent random variables, i.e. their joint PDF is equal to the
product of the one-mode PDF's corresponding to each individual
amplitude and phase,
$$
{\cal P}^{(N)} \{s, \xi \}  = \prod_{ l {\cal 2 B}_N} P^{(a)}_l (s_{l})
P^{(\psi)}_l (\xi_{l})
$$
\item The phase factors $\psi_l$ are uniformly
distributed on the unit circle in the complex plane, i.e.
for any mode $l$
$$
P^{(\psi)}_l (\xi_{l}) = 1/2\pi.
$$
\end{enumerate}
Note that RPA does not fix any shape of the amplitude PDF's
and, therefore, can deal with strongly non-Gaussian wavefields.
Such study of non-Gaussianity and intermittency of WT was
presented in \cite{ln,cln} and will not be repeated here.
However, we will study some new objects describing statistics of the phase.

In \cite{ln,cln}  RPA was {\em assumed} to hold over the nonlinear time.
The main goal of this paper is to find out whether it is true that
the RPA property survives over
the nonlinear time and to what extent. We will see that
RPA fails to hold in its pure form as formulated above but it survives
in the leading order  so that
the WT closure built using the RPA  is valid.
We will also see that independence of the the phase factors is
quite straightforward, whereas the amplitude independence is
subtle. Namely, $M$ amplitudes are independent only up to
a $O(M/N)$ correction. Based on this knowledge,
and leaving justification for later on in this paper,
we thus reformulate RPA 
in a weaker form which holds over the
nonlinear time and which involves $M$-mode PDF's with $M \ll N$ rather
than the full $N$-mode PDF.

\subsection{Definition of an essentially RPA field}

We will say that the field $a$ is of an ``essentially RPA'' type if:

\begin{enumerate}

\item The phase factors are statistically independent and uniformly
distributed variables up to  $O(\e^2)$ corrections, i.e.
\BE
{\cal P}^{(N)} \{s, \xi \}  = {1 \over (2 \pi)^{N} } {\cal P}^{(N,a)} \{s \} 
 \; [1 + O(\e^2)],
\EE
where 
\BE
 {\cal P}^{(N,a)} \{s \} = 
\left(
\prod_{ l {\cal 2 B}_N } 
\oint_{{\cal S}^{1} }
|d \xi_l| \; \right) {\cal P}^{(N)} \{s, \xi \},
\EE
is the $N$-mode {\em amplitude} PDF.

\item
The amplitude variables are almost independent is a sense
that for each $M \ll N$ modes the $M$-mode
amplitude PDF is equal to the product of
the one-mode PDF's up to $O(M/N)$ and $o(\e^2)$
corrections,
\BE
{\cal P}_{j_1, j_2, \dots , j_M} = 
 P^{(a)}_{j_1}  P^{(a)}_{j_2} \dots  P^{(a)}_{j_M} \; [1 +
O(M/N) + O(\e^2)].
\EE

\end{enumerate}

\section{  Weak nonlinearity and separation of time scales}

Consider weakly nonlinear dispersive waves in a periodic box. Here
we consider quadratic nonlinearity and the linear dispersion
relations $\omega_k$ which allow three-wave interactions. Example
of such systems include surface capillary waves~\cite{Zakfil,{ZakharovPRL}},
Rossby waves~\cite{bnaz}  and
internal waves in the ocean~\cite{LT}.  In Fourier space, we have
the following Hamiltonian equations,
\BEA i \, \dot a_l &=& \epsilon \sum_{m,n=1}^\infty \left(
V^l_{mn} a_{m} a_{n}e^{i\omega_{mn}^l t} \, \delta^l_{m+n}
 + 2 \bar{V}^{m}_{ln} \bar a_{n}
a_{m} e^{-i\omega^m_{ln}t } \, \delta^m_{l+n}\right),
\label{Interaction} \EEA
where $a_l=a(k_l)$ is the complex wave amplitude in the
interaction representation, $k_l = 2 \pi l/L $ is the wavevector, 
$L $ is the box side length, 
$\omega^l_{mn}\equiv\omega_{k_l}-\omega_{k_m}-\omega_{k_m}$,
$\omega_l=\omega_{k_l}$ is the wave frequency,
 $V^l_{mn}$ is an interaction coefficient and
$\epsilon$ is  a formal small nonlinearity parameter.

In order to filter out fast oscillations at the wave period, let
us seek for the solution at time $T$ such that $2 \pi / \omega \ll
T \ll 1/\omega \epsilon^2$.  The second condition ensures that $T$
is a lot less than the nonlinear evolution time.  Now let us use a
perturbation expansion in small $\epsilon$,
\BE a_l(T)=a_l^{(0)}+\epsilon a_l^{(1)}+\epsilon^2 a_l^{(2)}.
\label{Expansion} \EE
Substituting this expansion in (\ref{Interaction}) we get in the
zeroth order
$ a_l^{(0)}(T)=a_l(0)\label{definitionofa} $,
i.e. the zeroth order term is time independent. This corresponds
to the fact that the interaction representation wave amplitudes
are constant in the linear approximation.  For simplicity, we will
write $a^{(0)}_l(0)= a_l$, understanding that a quantity is taken
at $T=0$ if its time argument is not mentioned explicitly.  The
first order is given by
\BEA a^{(1)}_l (T) = -i \sum_{m,n=1}^\infty \left(   V^l_{mn} a_m
a_n \Delta^l_{mn} \delta^l_{m+n}
 + 2
\bar{V}^m_{ln}a_m\bar{a}_n \bar\Delta^m_{ln}\delta^m_{l+n}
\right), \label{FirstIterate} \EEA
where $ \Delta^l_{mn}=\int_0^T e^{i\omega^l_{mn}t}d t =
({e^{i\omega^l_{mn}T}-1})/{i \omega^l_{mn}}. \label{NewellsDelta}
$
Here we have taken into account that $a^{(0)}_l(T)= a_l$ and
$a^{(1)}_k (0)=0$. Iterating one more time we get

\BEA a_l^{(2)} (T)  &=& \sum_{m,n, \mu, \nu}^\infty \left[ 2
V^l_{mn} \left( -V^m_{\mu \nu}a_n a_\mu a_\nu E[\omega^l_{n \mu
\nu},\omega^l_{mn}] \delta^m_{\mu + \nu} -2 \bar V^\mu_{m \nu}a_n
a_\mu \bar a_\nu \bar E[\omega^{l \nu}_{n
\mu},\omega^l_{mn}]\delta^\mu_{m + \nu}\right) \delta^l_{m+n}
\right.\CR && \left. + 2 \bar V^m_{ln}
 \left(-V^m_{\mu \nu}\bar a_n a_\mu a_\nu E[\omega^{ln}_{\mu \nu},-\omega^m_{ln}]
\delta^m_{\mu + \nu} - 2 \bar V^\mu_{m \nu}\bar a_n a_\mu \bar
a_\nu E[-\omega^\mu_{n \nu l},-\omega^m_{l n}]  \delta^\mu_{m +
\nu} \right) \delta^m_{l+ n} \right. \CR && \left. + 2 \bar
V^m_{ln} \left( \bar V^n_{\mu \nu}a_m \bar a_\mu  \bar a_\nu
\delta^n_{\mu + \nu} E[-\omega^m_{l\nu\mu},-\omega^m_{ln}] + 2
V^\mu_{n \nu}a_m \bar a_\mu  a_\nu E[\omega^{\mu l}_{\nu m},
-\omega^m_{ln}]\delta^\mu_{n + \nu}\right)\delta^m_{l+n}
\right],\CR\label{SecondIterate} \EEA

\noindent where we used $a^{(2)}_k (0)=0$ and introduced
$E(x,y)=\int_0^T \Delta(x-y)e^{i y t} d t .$
%

\section{Evolution of the multi-mode PDF}

In this section we will apply the approach of \cite{ln,cln} 
to derive the  evolution equation for
the multi-mode PDF via introducing a generating functional,
performing a weak-nonlinearity expansion and statistical
averaging aided by a new graphical technique. 
We are going to demonstrate the phase independence
property.  This will also prepare us to answer the question
of the next section: to what extend
the amplitudes are going to remain  statistically independent?

\subsection{Generating functional.  }

Introduction of generating functionals often simplifies statistical
derivations but it can be defined differently to suit a particular
technique. For our problem, the most useful form of the generating
functional is
\BEA
Z^{(N)} \{\lambda, \mu \} =  
{ 1 \over (2 \pi)^{N}} \langle
\prod_{l \in {\cal B}_N }   \; e^{\lambda_l A_l^2} \psi_l^{\mu_l}
\rangle , \label{Z}
\EEA
where $\{\lambda, \mu \} \equiv \{\lambda_l, \mu_l ; l \in {\cal
B}_N\}$ is a set of parameters, $\lambda_l {\cal 2 R}$ and $\mu_l
{\cal 2 Z}$.  
%
%

\BE 
{\cal P}^{(N)}  \{s, \xi \} = 
 {1 \over (2 \pi)^{N}}  \sum_{\{\mu \}}   \langle
\prod_{l \in {\cal B}_N } \delta (s_l - A_l^2) \, \psi_l^{\mu_l}
\xi_l^{-\mu_l} \rangle \label{pdf-delta} 
\EE
where $\{\mu \} \equiv \{ \mu_l \in {\cal  Z};  {l \in {\cal B}_N }
\}$. This expression can be verified by considering mean of a
function $f \{A^2,\psi \}$ using the averaging rule (\ref{pdfn})
and expanding $f$ in the angular harmonics $\psi_l^m; \; m \in
{\cal Z}$ (basis functions on the unit circle),

\BE f\{A^2,\psi \} = \sum_{\{ m \} } g\{m, A\}
 \, \prod_{l \in {\cal B}_N } \psi_l^{m_l},
\EE
where $\{m\} \equiv \{ m_l \in {\cal  Z}; {l \in {\cal B}_N } \}$ are
indices enumerating the angular harmonics. Substituting this into
(\ref{pdfn}) with PDF given by  (\ref{pdf-delta}) and taking into
account that any nonzero power of $\xi_l$ will give zero after the
integration over the unit circle, one can see that LHS=RHS, i.e.
that (\ref{pdf-delta}) is correct. Now we can easily represent
(\ref{pdf-delta}) in terms of the generating functional,
\BE
 {\cal P}^{(N)}  \{s, \xi \} = 
  \hat {\cal L}_\lambda^{-1}
 \sum_{\{\mu \}}
 \left( Z^{(N)} \{\lambda, \mu\}
\, \prod_{l \in {\cal B}_N } \xi_l^{-\mu_l} \right) \label{jointpdf}
\EE
where $\hat {\cal L}_\lambda^{-1}$ stands for inverse the Laplace transform
with respect to all $\lambda_l$ parameters and $\{\mu \} \equiv \{
\mu_l \in {\cal Z}; {l \in {\cal B}_N } \}$ are the angular harmonics
indices.


By definition, in RPA fields all variables $A_l$ and $\psi_l$ are
statistically independent and $\psi_l$'s are uniformly distributed on
the unit circle. Such fields imply the following form of the
generating functional
\BE Z^{(N)} \{\lambda, \mu \}  =  Z^{(N,a)}  \{\lambda \}
\, \prod_{l \in {\cal B}_N } \delta(\mu_l),
\label{z-rpa} \EE
where
\BE 
 Z^{(N,a)}  \{\lambda \}
=\langle 
\prod_{l \in {\cal B}_N } 
e^{\lambda_l A_l^2} \rangle
= Z^{(N)} \{\lambda, \mu\}|_{\mu=0}
\EE
is an $N$-mode 
generating function for the  amplitude statistics.
Here, the Kronecker symbol $\delta(\mu_l)$ ensures
independence of the PDF from the phase factors $\psi_l$.
As a first step in validating the RPA property we will have  to prove 
that the generating functional
remains of  form (\ref{z-rpa}) up to $1/N$ and $O(\e^2)$ 
corrections over the nonlinear time
provided it has this form at $t=0$. 

\subsection{Asymptotic expansion of the generating functional.}

%
%
Let us first obtain an asymptotic weak-nonlinearity expansion
for the generating functional
$Z\{\lambda, \mu\}$ exploiting the separation of the linear and
nonlinear time scales. \footnote{ Hereafter we omit superscript ${(N)}$
in the $N$-mode objects if it does not lead to a confusion.}  To
do this, we have to calculate $Z$ at the intermediate time $t=T$ via
substituting into it $a_j(T)$ from  (\ref{Expansion})
 and retaining the terms up to $O(\e^2)$ only.
This calculation is given in the Appendix and the result of it is:
\BE
 Z\{\lambda, \mu, T\} =  X\{\lambda, \mu,T\} +  \bar X \{\lambda, - \mu,T\} 
\label{zx}
\EE
with 
\BE
X\{\lambda, \mu,T\} =  X(0) +  (2 \pi)^{2N} \left<\prod_{\|l\|<N}
e^{\lambda_l|a_l\zer|^2}[\e J_1 +\e^2(J_2 +J_3+J_4+J_5)] \right>_A + O(\e^4), 
\label{xt}
\EE
where
\BEA
J_1 &=&   \left<\prod_l \psi_l^{(0)\mu_l} 
\sum_j (\lambda_j
+\frac{\mu_j}{2|a_j\zer|^2})a_j\one\bar a_j\zer \right>_\psi, \label{j1} \\
J_2 &=&   {1 \over 2} \left<\prod_l \psi_l^{(0)\mu_l} 
\sum_j (\lambda_j+
\lambda_j^2|a_j\zer|^2-\frac{\mu_j^2}{2|a_j\zer|^2})|a_j\one|^2
 \right>_\psi, \label{j2} \\
J_3 &=&   \left<\prod_l \psi_l^{(0)\mu_l} 
 \sum_j 
(\lambda_j + \frac{\mu_j}{2|a_j\zer|^2})a_j\two\bar a_j\zer 
 \right>_\psi, \label{j3} \\
J_4 &=&   \left<\prod_l \psi_l^{(0)\mu_l} 
\sum_j 
\left[\frac{\lambda_j^2}{2}+\frac{\mu_j}{4|a_j\zer|^4}\left(\frac{\mu_j}{2}-1\right)+\frac{\lambda_j
\mu_j}{2|a_j\zer|^2} \right](a_j\one \bar a_j\zer)^2
 \right>_\psi, \label{j4} \\
J_5 &=&   {1 \over 2} \left<\prod_l \psi_l^{(0)\mu_l} 
\sum_{j \ne k}\lambda_j\lambda_k(a_j\one\bar a_j\zer +\bar a_j\one
a_j\zer)a_k\one\bar a_k\zer + (\lambda_j
+\frac{\mu_j}{4|a_j\zer|^2})\frac{\mu_k}{|a_k\zer|^2}(a_k\one\bar a_k\zer
-\bar a_k\one a_k\zer)a_j\one\bar a_j\zer
 \right>_\psi, \label{j5} 
\EEA
where $\left< \cdot \right>_A$ and $\left< \cdot \right>_\psi$ denote
the averaging over the initial amplitudes and initial phases
(which can be done independently).  
Our next step will be to calculate the above terms by substituting
into them the values of $a\one$ and $a\two$ from (\ref{FirstIterate})
and (\ref{SecondIterate}) respectively.

\subsection{Statistical averaging and graphs. }

Let us consider the initial fields $a_k(0) = a^{(0)}_k$ 
are essentially RPA as defined above. We will perform
averaging over the statistics of the initial fields
in order to obtain an evolution equations, first for $Z$ and
then for the multi-mode PDF. 
The ultimate goal of this exercise is to prove that
the wavefield remains of the essentially RPA type over the
nonlinear time.

Let us  introduce a graphical  classification of the above terms which
will allow us to simplify the statistical averaging and to understand
which terms are dominant. We will only consider here  contributions from
 $J_1$ and  $J_2$ which will allow us to understand the basic method.
Calculation of the rest of the terms, $J_3$, $J_4$ and  $J_5$, 
follows the same principles and can be found in Appendix 2.
First, 
 The linear in $\e$ terms are represented by $J_1$ which, upon using
(\ref{FirstIterate}), becomes
\BEA 
J_1
&=&\left<\prod_l \psi_l^{\mu_l}
\sum_{j,m,n}(\lambda_j +\frac{\mu_j}{2A_j^2})
\left(V_{mn}^j
a_m a_n\Delta_{mn}^j\delta_{m+n}^j +2\bar V_{jn}^m a_m\bar a_n
\bar\Delta_{jn}^m\delta_{j+n}^m\right)\bar
a_j\right>_\psi.
\label{firstineps}\EEA
Hereafter we omit, for brevity of notation, the super-script $\small
(0)$ because no other super-scripts will appear from now on.

Let us introduce some graphical notations for a simple classification
of different contributions to this and to other (more lengthy)
formulae that will follow. Combination $V_{mn}^j \delta_{m+n}^j$ will
be marked by a vertex joining three lines with in-coming $j$ and
out-coming $m$ and $n$ directions.  Complex conjugate $\bar V_{mn}^j
\delta_{m+n}^j$ will be drawn by the same vertex but with the opposite
in-coming and out-coming directions.  Presence of $a_j$ and $\bar a_j$
will be indicated by dashed lines pointing away and toward the vertex
respectively.~\footnote{This technique provides a useful
classification method but not a complete mathematical description of
the terms involved.} Thus, the two terms in formula (\ref{firstineps})
can be schematically represented as follows,

\

\vskip 1cm

\
\BEA
C_1=
\parbox{40mm} 
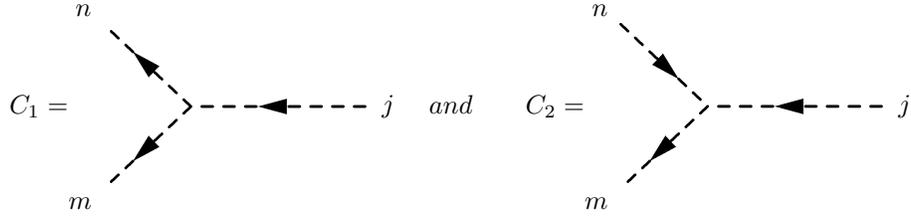
{
\begin{fmffile}{one}
   \begin{fmfgraph*}(110,62)
    \fmfleft{i1,i2}
    \fmfright{o1}
    \fmflabel{$m$}{i1}
    \fmflabel{$n$}{i2}
    \fmflabel{$j$}{o1}
    \fmf{dashes_arrow}{v1,i1}
    \fmf{dashes_arrow}{v1,i2}
 \fmf{dashes_arrow}{o1,v1}
   \end{fmfgraph*}
\end{fmffile}
}
&\quad  and \quad&
C_2=
\parbox{40mm} {
\begin{fmffile}{two}
   \begin{fmfgraph*}(110,62)
    \fmfleft{i1,i2}
    \fmfright{o1}
    \fmflabel{$m$}{i1}
    \fmflabel{$n$}{i2}
    \fmflabel{$j$}{o1}
    \fmf{dashes_arrow}{v1,i1}
    \fmf{dashes_arrow}{i2,v1}
 \fmf{dashes_arrow}{o1,v1}
   \end{fmfgraph*}
\end{fmffile}
}
\nonumber
\EEA
\vskip 1cm
Let us average over all the independent phase factors in the set
$\{\psi\}$.
Such averaging takes into account the statistical
independence and uniform distribution of
variables $\psi$. In particular, $\langle\psi\rangle=0$,
$\langle\psi_l \psi_m\rangle=0$ and $\langle\psi_l \bar
\psi_m\rangle=\delta^m_l$. Further, the products that involve
odd number of $\psi$'s are always zero, and among the even
products only those can survive that have equal numbers of
$\psi$'s and $\bar \psi$'s. These $\psi$'s and $\bar \psi$'s
must cancel each other which is possible if their indices
are matched in a pairwise way similarly to the  Wick's
theorem. The difference with the standard Wick, however, is
that there exists possibility of not only internal
(with respect to the sum) matchings but also external
ones with $\psi$'s in the pre-factor $\Pi \psi_l^{\mu_l}$.

Obviously, non-zero contributions can only arise for terms in which
all $\psi$'s cancel out either via internal mutual couplings within
the sum or via their external couplings to the $\psi$'s in the
$l$-product.  The internal couplings will indicate by joining the
dashed lines into loops whereas the external matching will be shown as
a dashed line pinned by a blob at the end. The number of blobs in
a particular graph will be called the {\em valence} of this graph.

Note that there will be no
contribution from the internal couplings between the incoming and the
out-coming lines of the same vertex because, due to the
$\delta$-symbol, one of the wavenumbers is 0 in this case, which means
\footnote{In the present paper we consider only spatially homogeneous wave
turbulence fields. In spatially homogeneous fields, due to momentum
conservation, there is no coupling to the zero mode $k=0$ because such
coupling would violate momentum conservation. Therefore if one of the
arguments of the interaction matrix element $V$ is equal to zero, the
matrix element is identically zero. That is to say that for any
spatially homogeneous wave turbulence system
$ V^{k=0}_{k_1 k_2} =  V^{k}_{k_1=0 k_2}  = V^{k}_{k_1 k_2=0} =0.$}
that $V=0$. For $J_1$ we have
$$ J_1 = \langle C_1 \rangle_\psi + \langle C_2 \rangle_\psi,$$ with
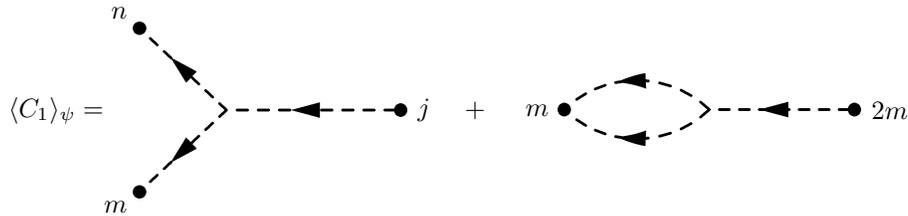
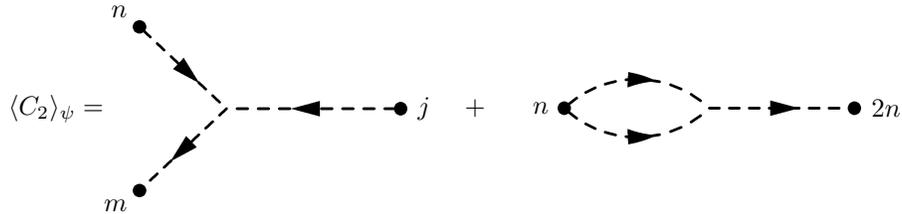
\BEA
 \langle C_1 \rangle_\psi  =
\parbox{40mm} {
\begin{fmffile}{three}
   \begin{fmfgraph*}(110,62)
    \fmfleft{i1,i2}
    \fmfright{o1}
    \fmflabel{$m$}{i1}
    \fmflabel{$n$}{i2}
    \fmflabel{$j$}{o1}
    \fmf{dashes_arrow}{v1,i1}
    \fmf{dashes_arrow}{v1,i2}
 \fmf{dashes_arrow}{o1,v1}
\fmfdot{i1}
\fmfdot{i2}
\fmfdot{o1}
   \end{fmfgraph*}
\end{fmffile}
}
&\quad  + \quad \quad&
\parbox{40mm} {
\begin{fmffile}{four}
   \begin{fmfgraph*}(110,62)
    \fmfleft{i1}
   \fmfforce{0.5w,0.5h}{v1}
    \fmfright{o1}
    \fmflabel{$m$}{i1}
    \fmflabel{$2m$}{o1}
    \fmf{dashes_arrow, left=.4, tension=.3}{v1,i1}
    \fmf{dashes_arrow, right=.4, tension=.3 }{v1,i1}
 \fmf{dashes_arrow}{o1,v1}
\fmfdot{i1}
\fmfdot{o1}
   \end{fmfgraph*}
\end{fmffile}
}
\nonumber
\EEA
and
\BEA
 \langle C_2 \rangle_\psi  =
\parbox{40mm} {
\begin{fmffile}{five}
   \begin{fmfgraph*}(110,62)
    \fmfleft{i1,i2}
    \fmfright{o1}
    \fmflabel{$m$}{i1}
    \fmflabel{$n$}{i2}
    \fmflabel{$j$}{o1}
    \fmf{dashes_arrow}{v1,i1}
    \fmf{dashes_arrow}{i2,v1}
 \fmf{dashes_arrow}{o1,v1}
\fmfdot{i1}
\fmfdot{i2}
\fmfdot{o1}
   \end{fmfgraph*}
\end{fmffile}
}
&\quad  + \quad \quad&
\parbox{40mm} {
\begin{fmffile}{six}
   \begin{fmfgraph*}(110,62)
    \fmfleft{i1}
   \fmfforce{0.5w,0.5h}{v1}
    \fmfright{o1}
    \fmflabel{$n$}{i1}
    \fmflabel{$2n$}{o1}
    \fmf{dashes_arrow, left=.4, tension=.3}{i1,v1}
    \fmf{dashes_arrow, right=.4, tension=.3 }{i1,v1}
 \fmf{dashes_arrow}{v1,o1}
\fmfdot{o1}
\fmfdot{i1}
   \end{fmfgraph*}
\end{fmffile}
}
\nonumber
\EEA
\vskip 1cm
which correspond to the following expressions,
\vskip 1cm
\BEA 
 \langle C_1 \rangle_\psi  
&=&
\sum_{j \ne m \ne n}(\lambda_j +\frac{\mu_j}{2A_j^2})
V_{mn}^j
A_m A_n A_j \Delta_{mn}^j\delta_{m+n}^j \delta(\mu_m +1)
\delta(\mu_n +1) \delta(\mu_j-1) 
 \prod_{l \ne j,m,n} \delta (\mu_l)
 \nonumber \\
&+& \sum_{m}(\lambda_{2m} +\frac{\mu_{2m}}{2A_{2m}^2})
V_{mm}^{2m}
A_m^2  A_{2m} \Delta_{mm}^{2m} \delta(\mu_m +2)
 \delta(\mu_{2m}-1) 
 \prod_{l \ne m,2m} \delta (\mu_l)
\label{haha1}
\EEA
and
\BEA 
 \langle C_2 \rangle_\psi  =
&=&
2\sum_{j \ne m \ne n}(\lambda_j +\frac{\mu_j}{2A_j^2})
\bar V_{jn}^m A_m A_n A_j
\bar\Delta_{jn}^m \delta_{j+n}^m
 \delta(\mu_m +1) \delta(\mu_n -1) \delta(\mu_j-1) 
 \prod_{l \ne j,m,n} \delta (\mu_l)
 \nonumber \\
&+&
2\sum_{n}(\lambda_n +\frac{\mu_n}{2A_n^2})
\bar V_{nn}^{2n} A_{2n} A_n^2
\bar\Delta_{nn}^{2n}
 \delta(\mu_{2n} +1) \delta(\mu_n -2) 
 \prod_{l \ne n,2n} \delta (\mu_l).
\label{haha2}\EEA
Because  of  the $\delta$-symbols  involving  $\mu$'s,  it takes  very
special combinations  of the arguments $\mu$  in $Z\{ \mu  \}$ for the
terms  in  the above  expressions  to  be  non-zero.  For  example,  a
particular term in  the first sum of (\ref{haha1})  may be non-zero if
two $\mu$'s in the set $\{ \mu  \}$ are equal to 1 whereas the rest of
them are 0. But in this case  there is only one other term in this sum
(corresponding to the  exchange of values of $n$ and  $j$) that may be
non-zero too.   In fact,  only utmost  two terms in  the both
(\ref{haha1})  and (\ref{haha2}) can  be non-zero  simultaneously.  In
the  other words,  each external  pinning of  the dashed  line removes
summation in  one index and, since  all the indices are  pinned in the
above diagrams, we are left with no summation at all in $J_1$ i.e. the
number of terms in $J_1$ is $O(1)$ with respect to large $N$.  We will
see   later   that    the   dominant   contributions   have   $O(N^2)$
terms. Although  these terms  come in the  $\e^2$ order, they  will be
much greater  that the $\e^1$ terms  because the limit  $N \to \infty$
must always be taken before $\e \to 0$.




Let us consider the first of the $\e^2$-terms, $J_2$. Substituting
 (\ref{FirstIterate}) into (\ref{j2}), we have
\vskip 1cm
\BEA J_2 &=& \frac{1}{2}\langle\prod_l\psi_l^{(0)\mu_l}
\sum_{j,m,n,\kappa,\nu}(\lambda_j+\lambda_j^2A_j^2-\frac{\mu_j^2}{2A_j^2})\CR 
&&\hspace{1cm} \times (V_{mn}^ja_ma_n\Delta_{mn}^j\delta_{m+n}^j+2\bar V_{jn}^ma_m\bar a_n\bar\Delta_{jn}^m\delta_{j+n}^m)
(\bar V_{\kappa\nu}^j\bar a_{\kappa}\bar a_{\nu}\bar\Delta_{\kappa\nu}^j\delta_{\kappa +\nu}^j
+2V_{j\nu}^{\kappa}\bar a_{\kappa}a_{\nu}\Delta_{j\nu}^{\kappa}\delta_{j+\nu}^{\kappa})\rangle_{\psi}\CR 
&=& \langle B_1 + B_2 + \bar B_2 + B_3 \rangle_{\psi}, \EEA
where
\\
\BEA
B_1 = \hspace{1cm}
\parbox{35mm} {
\begin{fmffile}{n7}
   \begin{fmfgraph*}(60,45)
\fmfforce{(0.w,1.h)}{i1}
\fmfforce{(0.w,0.h)}{i2}
\fmfforce{(1.w,1.h)}{o1}
\fmfforce{(1.w,0.h)}{o2}
\fmfforce{(0.25w,0.5h)}{v1}
\fmfforce{(0.75w,0.5h)}{v2}
\fmfforce{(0.5w,1.h)}{v3}
    \fmflabel{$m$}{i1}
    \fmflabel{$n$}{i2}
    \fmflabel{$\kappa$}{o1}
    \fmflabel{$\nu$}{o2}
    \fmf{dashes_arrow}{v1,i1}
    \fmf{dashes_arrow}{v1,i2}
    \fmf{dashes_arrow}{o1,v2}
    \fmf{dashes_arrow}{o2,v2}
\fmf{dots_arrow, label=$j$}{v2,v1}
   \end{fmfgraph*}
\end{fmffile}
}
&
B_2 = \hspace{1cm}
\parbox{35mm} {
\begin{fmffile}{n8}
   \begin{fmfgraph*}(60,45)
\fmfforce{(0.w,1.h)}{i1}
\fmfforce{(0.w,0.h)}{i2}
\fmfforce{(1.w,1.h)}{o1}
\fmfforce{(1.w,0.h)}{o2}
\fmfforce{(0.25w,0.5h)}{v1}
\fmfforce{(0.75w,0.5h)}{v2}
\fmfforce{(0.5w,1.h)}{v3}
    \fmflabel{$m$}{i1}
    \fmflabel{$n$}{i2}
    \fmflabel{$\kappa$}{o1}
    \fmflabel{$\nu$}{o2}
    \fmf{dashes_arrow}{v1,i1}
    \fmf{dashes_arrow}{v1,i2}
    \fmf{dashes_arrow}{o1,v2}
    \fmf{dashes_arrow}{v2,o2}
\fmf{dots_arrow, label=$j$}{v2,v1}
   \end{fmfgraph*}
\end{fmffile}
} and
&
B_3 = \hspace{1cm}
\parbox{35mm} {
\begin{fmffile}{n9}
   \begin{fmfgraph*}(60,45)
\fmfforce{(0.w,1.h)}{i1}
\fmfforce{(0.w,0.h)}{i2}
\fmfforce{(1.w,1.h)}{o1}
\fmfforce{(1.w,0.h)}{o2}
\fmfforce{(0.25w,0.5h)}{v1}
\fmfforce{(0.75w,0.5h)}{v2}
\fmfforce{(0.5w,1.h)}{v3}
    \fmflabel{$m$}{i1}
    \fmflabel{$n$}{i2}
    \fmflabel{$\kappa$}{o1}
    \fmflabel{$\nu$}{o2}
    \fmf{dashes_arrow}{v1,i1}
    \fmf{dashes_arrow}{i2,v1}
    \fmf{dashes_arrow}{o1,v2}
    \fmf{dashes_arrow}{v2,o2}
\fmf{dots_arrow, label=$j$}{v2,v1}
   \end{fmfgraph*}
\end{fmffile}
}
\EEA 
\\
\\
\vskip 1cm
Here the graphical notation for the interaction coefficients $V$ and 
the amplitude $a$ is the same as introduced in the previous section and
the dotted line with index $j$ indicates that there is a summation over $j$ 
but there is no amplitude $a_j$ in the corresponding expression.

Let us now perform the phase averaging which corresponds to the internal and external
couplings of the dashed lines. For $\langle B_1\rangle_{\psi}$ we have
\vskip 1cm
\BEA
\langle B_1\rangle_{\psi} =  \hspace{1cm}
\parbox{35mm} {
\begin{fmffile}{n10}
   \begin{fmfgraph*}(60,50) \fmfkeep{x}
\fmfforce{(0.w,1.h)}{i1}
\fmfforce{(0.w,0.h)}{i2}
\fmfforce{(1.w,1.h)}{o1}
\fmfforce{(1.w,0.h)}{o2}
\fmfforce{(0.25w,0.5h)}{v1}
\fmfforce{(0.75w,0.5h)}{v2}
\fmfforce{(0.5w,1.h)}{v3}
    \fmflabel{$m$}{i1}
    \fmflabel{$n$}{i2}
    \fmflabel{$\kappa$}{o1}
    \fmflabel{$\nu$}{o2}
    \fmf{dashes_arrow}{v1,i1}
    \fmf{dashes_arrow}{v1,i2}
    \fmf{dashes_arrow}{o1,v2}
    \fmf{dashes_arrow}{o2,v2}
\fmf{dots_arrow, label=$j$}{v2,v1}
\fmfdot{i1}
\fmfdot{i2}
\fmfdot{o1}
\fmfdot{o2}
   \end{fmfgraph*}
\end{fmffile}
} 
+
&
\parbox{35mm} {
\begin{fmffile}{n11}
   \begin{fmfgraph*}(70,50) \fmfkeep{fish}
\fmfforce{(0.w,1.h)}{i1}
\fmfforce{(0.w,0.h)}{i2}
\fmfforce{(1.w,.5h)}{o1}
\fmfforce{(0.2w,0.5h)}{v1}
\fmfforce{(0.6w,0.5h)}{v2}
    \fmflabel{$m$}{i1}
    \fmflabel{$n$}{i2}
    \fmf{dashes_arrow}{v1,i1}
    \fmf{dashes_arrow}{v1,i2}
    \fmf{dashes_arrow, left=1., label= $\nu$}{o1,v2}
    \fmf{dashes_arrow, right=1., label= $\nu$}{o1,v2}
\fmf{dots_arrow, label=$2 \nu$}{v2,v1}
\fmfdot{i1}
\fmfdot{i2}
\fmfdot{o1}
   \end{fmfgraph*}
\end{fmffile}
} 
+
&
\parbox{35mm} {
\begin{fmffile}{n12}
   \begin{fmfgraph*}(70,50) 
\fmfforce{(0.w,1.h)}{i1}
\fmfforce{(0.w,0.h)}{i2}
\fmfforce{(1.w,.5h)}{o1}
\fmfforce{(0.2w,0.5h)}{v1}
\fmfforce{(0.6w,0.5h)}{v2}
    \fmflabel{$\kappa$}{i1}
    \fmflabel{$\nu$}{i2}
    \fmf{dashes_arrow}{i1,v1}
    \fmf{dashes_arrow}{i2,v1}
    \fmf{dashes_arrow, left=1., label= $m$}{v2,o1}
    \fmf{dashes_arrow, right=1., label= $m$}{v2,o1}
\fmf{dots_arrow, label=$2 m$}{v1,v2}
\fmfdot{i1}
\fmfdot{i2}
\fmfdot{o1}
   \end{fmfgraph*}
\end{fmffile}
} \nonumber
\EEA
\BEA 
+2 \hspace{.5cm} 
\parbox{35mm} {
\begin{fmffile}{n13}
   \begin{fmfgraph*}(70,50) \fmfkeep{theta}
\fmfforce{(0.1w,0.5h)}{v1}
\fmfforce{(0.9w,0.5h)}{v2}
\fmf{dots_arrow, label=$j$}{v2,v1}
    \fmf{dashes_arrow, left=.7, label= $m$}{v1,v2}
    \fmf{dashes_arrow, right=.7, label= $n$}{v1,v2}
   \end{fmfgraph*}
\end{fmffile}
} ,
\label{B_1diagram}
\EEA
\\
where
\\
\BEA 
\parbox{25mm} {\fmfreuse{x}}
&=&
\frac{1}{2}\sum_{j\neq m\neq n\neq\kappa\neq\nu}(\lambda_j+\lambda_j^2A_j^2-\frac{\mu_j^2}{2A_j^2})
V_{mn}^j\bar V_{\kappa\nu}^j\Delta_{mn}^j\bar\Delta_{\kappa\nu}^j\delta_{m+n}^j\delta_{\kappa+\nu}^jA_mA_nA_{\kappa}A_{\nu}
\nonumber \\
&&  \hspace{2.2cm}  \times
\delta(\mu_m+1)\delta(\mu_n+1)\delta(\mu_{\kappa}-1)\delta(\mu_{\nu}-1)
\prod_{l\neq m,n,\kappa,\nu}\delta(\mu_l)
\nonumber
\EEA
\\
\\
\BEA 
\parbox{35mm} {\fmfreuse{fish}}
&=&\frac{1}{2}\sum_{m\neq n\neq\nu}(\lambda_{2\nu}+\lambda_{2\nu}^2A_{2\nu}^2-\frac{\mu_{2\nu}^2}{2A_{2\nu}^2})
V_{mn}^{2\nu}\delta_{m+n}^{2\nu}\bar V_{\kappa\nu}^{2\nu}
\nonumber\\ 
 &&\times
\Delta_{mn}^{2\nu}\bar\Delta_{\kappa\nu}^{2\nu}A_mA_nA_{\nu}^2
\delta(\mu_m+1)\delta(\mu_n+1)\delta(\mu_{\nu}-2)
\prod_{l\neq m,n,\nu}\delta(\mu_l)
\nonumber
\EEA
\\
\BEA 
2 \hspace{.5cm} \parbox{35mm} {\fmfreuse{theta}}
=\prod_{l}\delta(\mu_l)\sum_{j,m,n}(\lambda_{j}+\lambda_{j}^2A_{j}^2
)
|V_{mn}^{j}|^2
|\Delta_{mn}^{j}|^2
\delta_{m+n}^{j}A_m^2A_n^2
\nonumber
\EEA
\\
\\
\vskip 1cm
We have not written out the third term in (\ref{B_1diagram}) because
it is just a complex conjugate of the second one. Observe that all the
diagrams in the first line of (\ref{B_1diagram}) are $O(1)$ with
respect to large $N$ because all of the summations are lost due to the
external couplings(compare with the previous section). On the other
hand, the diagram in the second line contains two purely-internal
couplings and is therefore $O(N^2)$. This is because the number of
indices over which the summation survives is equal to the number of
purely internal couplings. Thus, the zero-valent graphs
are dominant and we can write
\BEA \langle B_1\rangle_\psi = \prod_l\delta(\mu_l)\sum_{j,m,n}(\lambda_{j}+\lambda_{j}^2A_{j}^2
)
|V_{mn}^{j}|^2
|\Delta_{mn}^{j}|^2
\delta_{m+n}^{j}A_m^2A_n^2[1+O(1/N^2)]\EEA

For $\langle B_2\rangle_\psi$ we have

\

\BEA
\langle B_2\rangle_{\psi} =  \hspace{1cm}
\parbox{35mm} {
\begin{fmffile}{n14}
   \begin{fmfgraph*}(60,50) \fmfkeep{x1}
\fmfforce{(0.w,1.h)}{i1}
\fmfforce{(0.w,0.h)}{i2}
\fmfforce{(1.w,1.h)}{o1}
\fmfforce{(1.w,0.h)}{o2}
\fmfforce{(0.25w,0.5h)}{v1}
\fmfforce{(0.75w,0.5h)}{v2}
    \fmflabel{$m$}{i1}
    \fmflabel{$n$}{i2}
    \fmflabel{$\kappa$}{o1}
    \fmflabel{$\nu$}{o2}
    \fmf{dashes_arrow}{v1,i1}
    \fmf{dashes_arrow}{v1,i2}
    \fmf{dashes_arrow}{o1,v2}
    \fmf{dashes_arrow}{v2,o2}
\fmf{dots_arrow, label=$j$}{v2,v1}
\fmfdot{i1}
\fmfdot{i2}
\fmfdot{o1}
\fmfdot{o2}
   \end{fmfgraph*}
\end{fmffile}
} 
+2 \hspace{1cm}
\parbox{35mm} {
\begin{fmffile}{n15}
   \begin{fmfgraph*}(60,50) \fmfkeep{hammock}
\fmfforce{(0.w,1.h)}{i1}
\fmfforce{(1.w,1.h)}{o1}
\fmfforce{(0.25w,0.5h)}{v1}
\fmfforce{(0.75w,0.5h)}{v2}
    \fmflabel{$m$}{i1}
    \fmflabel{$-m$}{o1}
    \fmf{dashes_arrow}{v1,i1}
    \fmf{dashes_arrow}{v2,o1}
\fmf{dots_arrow, label=$j$}{v2,v1}
\fmf{dashes_arrow, right=.7, label=$n$}{v1,v2}
\fmfdot{i1}
\fmfdot{o1}
   \end{fmfgraph*}
\end{fmffile}
} 
+2 \hspace{1cm}
\parbox{35mm} {
\begin{fmffile}{n16}
   \begin{fmfgraph*}(60,50) \fmfkeep{A}
\fmfforce{(0.w,0.h)}{i1}
\fmfforce{(1.w,0.h)}{o1}
\fmfforce{(0.25w,0.5h)}{v1}
\fmfforce{(0.75w,0.5h)}{v2}
\fmfforce{(0.5w,1.h)}{v3}
   \fmflabel{$n$}{v3}
    \fmflabel{$m$}{i1}
    \fmflabel{$\kappa$}{o1}
    \fmf{dashes_arrow}{v1,i1}
    \fmf{dashes_arrow}{o1,v2}
\fmf{dots_arrow, label=$j$}{v2,v1}
\fmf{dashes_arrow}{v1,v3}
\fmf{dashes_arrow}{v2,v3}
\fmfdot{i1}
\fmfdot{o1}
\fmfdot{v3}
   \end{fmfgraph*}
\end{fmffile}
} 
\nonumber
\EEA
\\
\BEA
+ \hspace{1cm}
\parbox{35mm} {
\begin{fmffile}{n17}
   \begin{fmfgraph*}(70,50) \fmfkeep{fish1}
 \fmfforce{(0.w,1.h)}{i1}
\fmfforce{(0.w,0.h)}{i2}
\fmfforce{(1.w,.5h)}{o1}
\fmfforce{(0.2w,0.5h)}{v1}
\fmfforce{(0.6w,0.5h)}{v2}
    \fmflabel{$\kappa$}{i1}
    \fmflabel{$\nu$}{i2}
    \fmf{dashes_arrow}{i1,v1}
    \fmf{dashes_arrow}{v1,i2}
    \fmf{dashes_arrow, left=1., label= $m$}{v2,o1}
    \fmf{dashes_arrow, right=1., label= $m$}{v2,o1}
\fmf{dots_arrow, label=$2 m$}{v1,v2}
\fmfdot{i1}
\fmfdot{i2}
\fmfdot{o1}
   \end{fmfgraph*}
\end{fmffile}
} 
+ \hspace{1cm}
\parbox{45mm} {
\begin{fmffile}{n18}
   \begin{fmfgraph*}(85,50) \fmfkeep{scorpion1}
 \fmfforce{(0.w,.5h)}{i1}
\fmfforce{(1.w,.5h)}{o1}
\fmfforce{(0.25w,0.5h)}{v1}
\fmfforce{(0.6w,0.5h)}{v2}
    \fmflabel{$3m$}{i1}
     \fmf{dashes_arrow}{i1,v1}
    \fmf{dashes_arrow, left=.7, label= $m$}{v2,o1}
    \fmf{dashes_arrow, right=.7, label= $m$}{v2,o1}
\fmf{dots_arrow, label=$2 m$}{v1,v2}
    \fmf{dashes_arrow, left=1., label= $m$}{v1,o1}
\fmfdot{i1}
\fmfdot{o1}
   \end{fmfgraph*}
\end{fmffile}
} 
+ \hspace{1cm}
\parbox{45mm} {
\begin{fmffile}{n19}
   \begin{fmfgraph*}(85,50) \fmfkeep{scorpion2}
 \fmfforce{(0.w,.5h)}{i1}
\fmfforce{(1.w,.5h)}{o1}
\fmfforce{(0.25w,0.5h)}{v1}
\fmfforce{(0.6w,0.5h)}{v2}
    \fmflabel{$-m$}{i1}
     \fmf{dashes_arrow}{v1,i1}
    \fmf{dashes_arrow, left=.7, label= $m$}{v2,o1}
    \fmf{dashes_arrow, right=.7, label= $m$}{v2,o1}
\fmf{dots_arrow, label=$2 m$}{v1,v2}
    \fmf{dashes_arrow, right=1., label= $m$}{o1,v1}
\fmfdot{i1}
\fmfdot{o1}
   \end{fmfgraph*}
\end{fmffile}
} 
\label{b2}
\EEA
\\
where
\\
\BEA \parbox{25mm} {\fmfreuse{x1}} 
&=&
\sum_{j\neq m\neq n\neq\kappa\neq\nu}(\lambda_j+\lambda_j^2A_j^2-\frac{\mu_j^2}{2A_j^2})
V_{mn}^j V_{j\nu}^{\kappa}
\Delta_{mn}^j\Delta_{j\nu}^{\kappa}\delta_{m+n}^j\delta_{j+\nu}^{\kappa}A_mA_nA_{\kappa}A_{\nu}
\nonumber \\
&& \hspace{2.2cm} \times
\delta(\mu_m+1)\delta(\mu_n+1)\delta(\mu_{\kappa}-1)\delta(\mu_{\nu}+1)
\prod_{l\neq m,n,\kappa,\nu}\delta(\mu_l)
\nonumber
\EEA
\\
\\
\BEA \parbox{35mm} {\fmfreuse{hammock}} 
&=& \sum_{j,m,n}(\lambda_j+\lambda_j^2A_j^2-\frac{\mu_j^2}{2A_j^2})
V_{mn}^j V_{j-m}^{n}\Delta_{mn}^j 
\nonumber \\ \hskip 5cm && \times 
\Delta_{j-m}^{n}\delta_{m+n}^jA_n^2A_mA_{-m}
\delta(\mu_m+1)\delta(\mu_{-m}+1)
\prod_{l\neq m,-m}\delta(\mu_l)
\nonumber
\EEA
\\
 \\
\BEA \parbox{25mm} {\fmfreuse{A}}  &=&
\sum_{j\neq m\neq n\neq\kappa}(\lambda_j+\lambda_j^2A_j^2-\frac{\mu_j^2}{2A_j^2})
V_{mn}^j V_{jn}^{\kappa}
\nonumber \\ \hskip 5cm && \times 
\Delta_{mn}^j \Delta_{jn}^{\kappa}\delta_{m+n}^j\delta_{j+n}^{\kappa}A_mA_n^2A_{\kappa}
\delta(\mu_m+1)\delta(\mu_n+2)\delta(\mu_{\kappa}-1)
\prod_{l\neq m,n,\kappa}\delta(\mu_l)
\nonumber
\EEA
\\
\\
\BEA \parbox{35mm} {\fmfreuse{fish1}} 
&=&
\sum_{ m\neq \kappa\neq\nu}(\lambda_{2m}+\lambda_{2m}^2A_{2m}^2-\frac{\mu_{2m}^2}{2A_{2m}^2})
V_{mm}^{2m} V_{2m\nu}^{\kappa}
\nonumber \\ \hskip 5cm && \times 
\Delta_{mm}^{2m} \Delta_{2m\nu}^{\kappa}\delta_{2m+\nu}^{\kappa}A_m^2A_{\kappa}A_{\nu}
\hspace{1.2cm}
\nonumber \\
&& \hspace{2.2cm} \times
\delta(\mu_m+2)\delta(\mu_{\kappa}-1)\delta(\mu_{\nu}+1)
\prod_{l\neq m,\kappa,\nu}\delta(\mu_l)
\nonumber
\EEA
\BEA \parbox{35mm} {\fmfreuse{scorpion1}} 
&=&
\sum_{ m}(\lambda_{2m}+\lambda_{2m}^2A_{2m}^2-\frac{\mu_{2m}^2}{2A_{2m}^2})
V_{mm}^{2m} V_{2mm}^{3m}
\nonumber \\ \hskip 5cm && \times 
\Delta_{mm}^{2m} \Delta_{2mm}^{3m} A_m^3 A_{3m}
\delta(\mu_m+3)\delta(\mu_{3m}-1)
\prod_{l\neq m, 3m}\delta(\mu_l)
\nonumber
\EEA
\\
\BEA \parbox{35mm} {\fmfreuse{scorpion2}} 
&=&
\sum_{ m}(\lambda_{2m}+\lambda_{2m}^2A_{2m}^2-\frac{\mu_{2m}^2}{2A_{2m}^2})
V_{mm}^{2m} V_{2m -m}^{m}\Delta_{mm}^{2m}
\nonumber \\ \hskip 5cm && \times 
\Delta_{2m \, -m}^{m} A_m^3 A_{-m}
\delta(\mu_m+1) \delta(\mu_{-m}+1)
\prod_{l\neq m, -m}\delta(\mu_l)
\nonumber
\EEA
The second term in (\ref{b2}) contains one summation because its graph
has one purely internal coupling. This term is $N$ times smaller than
the largest terms in $\langle B_1 \rangle_\psi$ (which have 2
surviving summation indices).  All the other terms in (\ref{b2})
contain no summation at all because all their dashed lines are coupled
externally.

Similarly, the leading contribution to $\langle B_3 \rangle_\psi$ will
be given by the zero-valent graph with the maximum possible number of internal
couplings (which is equal to 2 in this case). Because of the
$\delta$'s, there are no graphs with just one internal coupling, but
there are graphs with all the dashed lines coupled externally.  Thus,
\vskip 1cm
\BEA
\langle B_3 \rangle_\psi &=&
\parbox{25mm} {
\begin{fmffile}{n19p}
   \begin{fmfgraph*}(70,50) \fmfkeep{theta1}
\fmfforce{(0.1w,0.5h)}{v1}
\fmfforce{(0.9w,0.5h)}{v2}
\fmf{dots_arrow, label=$j$}{v2,v1}
    \fmf{dashes_arrow, left=.7, label= $m$}{v1,v2}
    \fmf{dashes_arrow, left=.7, label= $n$}{v2,v1}
   \end{fmfgraph*}
\end{fmffile}
} 
[1+O(1/N^2)]
\nonumber\\ && 
=2 \prod_{l}\delta(\mu_l)\sum_{j,m,n}(\lambda_{j}+\lambda_{j}^2A_{j}^2
)
|V_{jn}^{m}|^2 |\Delta_{jn}^{m}|^2 \delta_{j+n}^{m}A_m^2A_n^2 \; [1+O(1/N^2)]
,
\label{b3}
\EEA
Summarising the results of this section we can write for $J_2$:
\BEA
J_2 = \prod_{l}\delta(\mu_l)\sum_{j,m,n}(\lambda_{j}+\lambda_{j}^2A_{j}^2
)
\left[|V_{mn}^{j}|^2 |\Delta_{mn}^{j}|^2 \delta_{m+n}^{j} +
2 |V_{jn}^{m}|^2 |\Delta_{jn}^{m}|^2 \delta_{j+n}^{m}
\right]
A_m^2A_n^2 \; [1+O(1/N)].
\EEA
Thus, we considered in detail the different terms involved
in $J_2$ and we found that the dominant contributions come
from the zero-valent graphs because the have more summation
indices involved. This turns out to be the general rule that
allows one to simplify calculation by discarding a significant
number of graphs with non-zero valence.
After this observation finding  the rest of the terms,
$J_3$ to $J_5$, becomes a routine task and we therefore move it to
the Appendix 2.


\subsection{Equation for $Z$ \label{EQZ}}


Now we can observe that all contributions to the
evolution of $Z$ (namely $J_1 - J_5$, see the previous section and Appendix 2)
 contain factor $
\prod_{l}\delta(\mu_l) $ which means that the phase factors $\{\psi
\}$ remain a set of statistically independent (of each each other and
of $A$'s) variables uniformly distributed on $S^1$. This is true with
accuracy $O(\e^2)$ (assuming that the $N$-limit is taken first, i.e.
$1/N \ll \e^2$) and this proves persistence of the first of the
``essential RPA'' properties. Similar result
for a special class of three-wave systems arising in the solid state physics
was previously obtained by Brout and Prigogine \cite{bp}.
This result is interesting 
because it has been obtained without any assumptions on the
statistics of the amplitudes $\{ A \}$ and, therefore, it is valid
beyond the RPA approach. It may appear useful in future for study of
fields with random phases but correlated amplitudes.

Let us now derive an evolution equation for the generating functional.
Using our results for $J_1 - J_5$ in (\ref{xt}) and (\ref{zx}) we have
\BEA
Z(T) - Z(0) &=& \e^2
 \sum_{j,m,n}(\lambda_{j}+\lambda_{j}^2 {\partial \over \partial \lambda_{j}})
\left[|V_{mn}^{j}|^2 |\Delta_{mn}^{j}|^2 \delta_{m+n}^{j} +
2 |V_{jn}^{m}|^2 |\Delta_{jn}^{m}|^2 \delta_{j+n}^{m}
\right]
{\partial^2 Z(0)\over \partial \lambda_{m} \partial \lambda_{n}} 
\nonumber \\
&& + 4\e^2 \sum_{j,m,n}
\lambda_j
\left[ - |V_{mn}^j|^2\bar E(0,\omega_{mn}^j)\delta_{m+n}^j
{\partial \over \partial \lambda_{n}}
+|V_{jn}^m|^2 E(0,-\omega_{jn}^m)\delta_{j+n}^m 
\left(
{\partial \over \partial \lambda_{m}}
- {\partial \over \partial \lambda_{n}} \right)
\right]  
{\partial Z(0) \over \partial \lambda_{j}}
\nonumber \\
&& + 2 \e^2 \sum_{j\neq k,n}\lambda_j\lambda_k
\left[ 
-2 |V_{kn}^j|^2 \delta_{k+n}^j |\Delta_{kn}^j|^2 
+|V_{jk}^n|^2 \delta_{j+k}^n |\Delta_{jk}^n|^2 
\right]
{\partial^3 Z(0)\over \partial \lambda_{j} \partial \lambda_{n}
\partial \lambda_{k}} \; +cc.
\label{discreteZ}
\EEA
Here partial derivatives with respect to $\lambda_l$ appeared because of 
the $A_l$ factors.
This expression is valid up to $O(\e^4)$
and $O(\epsilon^2/N)$ corrections.
Note that we still have not used any assumption about the statistics
of $A$'s. This is a linear equation: as usual in statistics we traded
nonlinearity for higher dimensions.
The last term here ``spoils''  the separation of variables and,
therefore, puts a question mark on the independence of
variables $\{A \}$ from each other on the nonlinear time.

Let us now  $N \to \infty$ limit followed by $T \sim 1/\epsilon \to \infty$
(we re-iterate that this order of the limits is essential).
Taking into account that
$\lim\limits_{T\to\infty}E(0,x)= T (\pi
\delta(x)+iP(\frac{1}{x}))$, and
$\lim\limits_{T\to\infty}|\Delta(x)|^2=2\pi T\delta(x)$ and,
replacing $(Z(T) -Z(0))/T$ by $\dot Z$ we have
\BEA
\dot Z	&=&
4 \pi \e^2
 \int \big\{ (\lambda_{j}+\lambda_{j}^2 {\delta \over \delta \lambda_{j}})
\left[|V_{mn}^{j}|^2 \delta(\omega_{mn}^{j}) \delta_{m+n}^{j} +
2 |V_{jn}^{m}|^2 \delta(\omega_{jn}^{m}) \delta_{j+n}^{m}
\right]
{\delta^2 Z\over \delta \lambda_{m} \delta \lambda_{n}} 
\nonumber \\
&& + 2
\lambda_j
\left[ - |V_{mn}^j|^2 \delta(\omega_{mn}^j)\delta_{m+n}^j
{\delta \over \delta \lambda_{n}}
+|V_{jn}^m|^2 \delta(\omega_{jn}^m)\delta_{j+n}^m 
\left(
{\delta \over \delta \lambda_{m}}
- {\delta \over \delta \lambda_{n}} \right)
\right]  
{\delta Z \over \delta \lambda_{j}}
\nonumber \\
&& + 
2 \lambda_j\lambda_m
\left[ 
-2 |V_{mn}^j|^2 \delta_{m+n}^j \delta(\omega_{mn}^j)
+|V_{jm}^n|^2 \delta_{j+m}^n \delta(\omega_{jm}^n) 
\right]
{\delta^3 Z \over \delta \lambda_{j} \delta \lambda_{n}
\delta \lambda_{m}} \big\}\, dk_j dk_m dk_n.
\label{Zequat}
\EEA
Here variational derivatives appeared instead of partial derivatives because of 
the $N\to\infty$ limit.
%


\subsection{Equation for the PDF}


%
Taking the inverse Laplace transform of (\ref{Zequat}) we have the following
equation for the PDF,
\BE
\dot {\cal P} = - \int {\delta F_j \over \delta s_j} \, dk_j,
\label{peierls}
\EE
where $F_j$ is a flux of probability in the space of the amplitude $s_j$,
\BEA
F_j &=&
4 \pi \e^2 \int
\big\{ 
(|V_{mn}^{j}|^2 \delta(\omega_{mn}^{j}) \delta_{m+n}^{j} +
2 |V_{jm}^{n}|^2 \delta(\omega_{jm}^{n}) \delta_{j+m}^{n}
)
\left[ s_n s_m {\cal P} - {\delta \over \delta s_j} (s_j s_n s_m {\cal P}) \right]
\nonumber \\
&&
-2 {\cal P} \left[|V_{mn}^{j}|^2 \delta(\omega_{mn}^{j}) \delta_{m+n}^{j} s_j s_m
+ |V_{jm}^{n}|^2 \delta(\omega_{jm}^{n}) \delta_{j+m}^{n}
(s_j s_m - s_j s_n) \right] 
\nonumber \\
&&
-2
(|V_{jm}^{n}|^2 \delta(\omega_{jm}^{n}) \delta_{j+m}^{n}
-2|V_{mn}^{j}|^2 \delta(\omega_{mn}^{j}) \delta_{m+n}^{j} )
{\delta \over \delta s_m} (s_j s_n s_m {\cal P}) \big\} \, dk_m dk_n.
\EEA
This expression can be simplified to
\BEA
-{F_j \over
4 \pi \e^2 s_j} 
&=&
\int
\big\{ 
(|V_{mn}^{j}|^2 \delta(\omega_{mn}^{j}) \delta_{m+n}^{j} +
2 |V_{jm}^{n}|^2 \delta(\omega_{jm}^{n}) \delta_{j+m}^{n}
)
s_n s_m {\delta {\cal P} \over \delta s_j}
\nonumber \\
&&
+2 {\cal P} (
|V_{jm}^{n}|^2 \delta(\omega_{jm}^{n}) \delta_{j+m}^{n}
- |V_{mn}^{j}|^2 \delta(\omega_{mn}^{j}) \delta_{m+n}^{j} 
)s_m
\nonumber \\
&&
+2
(|V_{jm}^{n}|^2 \delta(\omega_{jm}^{n}) \delta_{j+m}^{n}
-2|V_{mn}^{j}|^2 \delta(\omega_{mn}^{j}) \delta_{m+n}^{j} )
s_n s_m
{\delta {\cal P} \over \delta s_m}
 \big\} \, dk_m dk_n.
\label{flux}
\EEA
This equation is identical to the one originally obtained
by Peierls \cite{peierls} and later rediscovered
by Brout and Prigogine \cite{bp} in the context of
the physics of anharmonic crystals.
Zaslavski and Sagdeev  \cite{zs} were the first to study this
equation in the WT context. However, the  analysis of 
\cite{peierls,bp,zs}
was restricted 
to the interaction Hamiltonians of the ``potential energy'' type,
i.e. the ones that involve only the coordinates but not the momenta.
This restriction leaves aside a great many important WT systems,
e.g. the capillary, Rossby, internal and MHD waves.
Our result above
indicates that the Peierls equation
 is also valid in the most general case of 3-wave systems.

Here we should again emphasise importance of the taken order
of limits, $N \to \infty$ first and $\e \to 0$ second.
Physically this means that the frequency resonance is broad
enough to cover great many modes. Some authors, e.g. \cite{peierls,bp,zs},
leave the sum notation in the PDF equation even after the
$\e \to 0$ limit taken giving $\delta(\omega_{jm}^{n})$.
One has to be careful interpreting such formulae because
formally the RHS is nill in most of the cases because 
there may be no exact resonances between the discrete $k$ modes
(as it is the case, e.g. for the capillary waves). In real finite-size
physical systems, this condition means that the wave amplitudes, although
small, should not be too small so that the frequency broadening is sufficient
to allow the resonant interactions. Our 
functional integral notation is meant to indicate that 
the $N \to \infty$ limit has already been taken.

\section{Approximate  independence of the amplitudes.}

Variables $s_j$ do not separate 
in the above equation for the PDF.
Indeed, substituting 
\BE
{\cal P}^{(N,a)} =
 P^{(a)}_{j_1}  P^{(a)}_{j_2} \dots  P^{(a)}_{j_N} \; 
\label{pure}
\EE
into the discrete version of (\ref{flux}) we
see that it turns into zero on the thermodynamic solution
with $P^{(a)}_{j} = \omega_j \exp(-\omega_j s_j)$.
However, it is not
zero for the one-mode PDF  $P^{(a)}_{j}$
corresponding to the cascade-type Kolmogorov-Zakharov (KZ)
spectrum $n_j^{kz}$, i.e.
 $P^{(a)}_{j} = (1/n_j^{kz}) \exp(-s_j /n_j^{kz})$ (see next section), 
nor it is likely to be zero for any other
PDF of form (\ref{pure}).
This means that, even initially independent, 
the amplitudes will correlate with each other
at the nonlinear time. Does this mean that the existing 
WT theory, and in particular the kinetic equation, is 
invalid?

To answer to this question let us differentiate the discrete version
of the equation (\ref{Zequat}) with respect to $\lambda$'s to get
equations for the amplitude moments. We can easily see that
\BE
\partial_t \left(\langle A_{j_1}^2 A_{j_2}^2 \rangle 
-\langle A_{j_1}^2 \rangle  \langle A_{j_2}^2 \rangle \right) =
O(\e^4) \quad (j_1, j_2 \in {\cal B}_N) 
\label{split}
\EE  
if $\langle A_{j_1}^2 A_{j_2}^2 A_{j_3}^2 \rangle =
\langle A_{j_1}^2 \rangle  \langle A_{j_2}^2 \rangle  \langle
A_{j_3}^2 \rangle $ (with the same accuracy)
at $t=0$.
Similarly, in terms of PDF's 
\BE
\partial_t \left(P^{(2,a)}_{j_1, j_2} (s_{j_1}, s_{j_2}) 
- P^{(a)}_{j_1}(s_{j_1}) P^{(a)}_{j_2}(s_{j_2})  \right) =
 O(\e^4) \quad (j_1, j_2  \in {\cal B}_N)
\EE  
if $P^{(4,a)}_{j_1, j_2, j_3, j_4} (s_{j_1}, s_{j_2}, s_{j_3}, s_{j_4}) =
P^{(a)}_{j_1}(s_{j_1}) P^{(a)}_{j_2}(s_{j_2}) P^{(a)}_{j_3}(s_{j_3})  P^{(a)}_{j_4}(s_{j_4}) 
$ at $t=0$.
Here $P^{(4,a)}_{j_1, j_2, j_3, j_4} (s_{j_1}, s_{j_2}, s_{j_3}, s_{j_4}) $,
$P^{(2,a)}_{j_1, j_2} (s_{j_1}, s_{j_2}) $ and 
$P^{(a)}_{j} (s_{j}) $ 
are
the four-mode,
two-mode and one-mode PDF's obtained
from $\cal P$ by integrating out all but 3,2 or  1
arguments respectively.
One can see that, with a $\e^2$ accuracy,
the Fourier modes will remain independent of each other
in any pair over the nonlinear time if they were
independent in every triplet at $t=0$.

Similarly, one can show that the modes will remain
independent over the nonlinear time
in any subset of $M<N$ modes with accuracy $M/N$
(and $\e^2$) if they were initially independent in
every subset of size $M+1$. Namely
\BEA
P^{(M,a)}_{j_1, j_2, \dots , j_M} (s_{j_1}, s_{j_2}, s_{j_M}) 
- P^{(a)}_{j_1}(s_{j_1}) P^{(a)}_{j_2}(s_{j_2} ) \dots
P^{(a)}_{j_M}(s_{j_M} )   =
O(M/N) + O(\e^2) \nonumber \\
\quad (j_1, j_2, \dots, j_M \in {\cal B}_N)
\EEA  
if $P^{(M+1,a)}_{j_1, j_2, \dots, j_{M+1}} =
P^{(a)}_{j_1}  P^{(a)}_{j_2} \dots P^{(a)}_{j_{M+1}}
$ at $t=0$.

Mismatch $O(M/N)$ arises from some terms in the ZS equation with
coinciding indices $j$. For $M=2$ there is only one such term
in the $N$-sum and, therefore, the corresponding error is
$O(1/N)$ which is much less than $O(\epsilon^2)$ (due to the
order of the limits in $N$ and $\epsilon$).
However, the number of such terms grows as $M$ and the error
accumulates to  $O( M/N)$ which can greatly exceed $O(\epsilon^2)$
for sufficiently large $M$.

We see that the accuracy with which the modes remain
independent in a subset is worse for larger
subsets and that the independence property is
completely lost for subsets  approaching in size
the entire set, $M \sim N$.
One should not worry too much about this loss
because $N$ is the biggest parameter in the 
problem (size of the box) and the modes
will be independent in all
$M$-subsets no matter how large.
Thus, the statistical objects
involving any {\em finite} number of modes
are factorisable as products of the one-mode
objects and, therefore, the WT theory  reduces to 
considering the one-mode objects. 
This results explains why we re-defined RPA in its
relaxed ``essential RPA'' form.
Indeed, in this form RPA is sufficient for the WT
closure and, on the other hand, it remains valid over the nonlinear
time. In particular, only property (\ref{split}) is needed,
as far as the amplitude statistics is concerned, for deriving
the 3-wave kinetic equation, and this fact validates this equation
and all of its solutions, including the KZ spectrum which plays an
important role in WT. 

The situation were modes can be considered as independent when
taken in relatively small sets but should be treated as
 dependent in the context
of much  larger sets is not so unusual in physics. Consider for
example a distribution of electrons and ions in plasma.
The full $N$-particle distribution function in this case satisfies
the Liouville  equation which is, in general, not a separable
equation. In other words, the $N$-particle distribution function
cannot be written as a product of $N$ 
one-particle distribution functions. However, an $M$-particle
distribution can indeed be represented as 
a product of $M$ 
one-particle distributions if $M \ll N_D$ where $N_D$ is the number
of particles in the Debye sphere. We see  an interesting transition
from a an individual to collective behaviour when the number
of particles  approaches $N_D$. In the special case of the one-particle
function we have here the famous mean-field Vlasov equation which is
valid up to  $O(1/N_D)$ corrections (representing particle collisions).

\section{One-mode statistics}

We have established above that 
the one-point statistics is at the heart of the WT theory.
All one-point statistical objects can be derived from the one-point
amplitude generating function, 
$$Z_a (\lambda_j) = \left< e^{\lambda_j A_j^2} \right> $$ 
which can be obtained from the $N$-point $Z$ by taking all $\mu$'s
and all $\lambda$'s, except for $\lambda_j$, equal to zero.
Substituting such values to (\ref{Zequat}) we get the following
equation for $Z_a$,
\begin{equation}
\frac{\partial Z_a}{\partial t} = \lambda_j \eta_j Z_a +(\lambda_j^2
\eta_j - \lambda_j \gamma_j) \frac{\partial Z_a}{\partial \lambda_j},
\label{za}
\end{equation}
where,
\BEA \eta_j  = 4 \pi \epsilon^2 \int 
\left(|V^j_{lm}|^2 \delta^j_{lm}  \delta(\omega^j_{lm})
+2 |V^m_{jl}|^2 \delta^m_{jl}  \delta(\omega^m_{jl} )
\right)  n_{l} n_{m}
\, d { k_l} d { k_m} ,  \label{RHO} \\
 \gamma_j =
8 \pi \epsilon^2 \int  
\left(
|V^j_{lm}|^2 \delta^j_{lm} \delta(\omega^j_{lm}) n_{m}
 +|V^m_{jl} |^2 \delta^m_{jl} \delta(\omega^m_{jl}) (n_{l}- n_{m})
\right) \, d { k_l} d { k_m}  .  \label{GAMMA}
\EEA
Correspondingly, for the one mode PDF $P_a (s_j) $ we have
\begin{equation}
{\partial P_a \over \partial t}+  {\partial F \over \partial s_j}  =0,
 \label{pa}
\end{equation}
 with $F$ is a probability flux in the s-space,
\begin{equation}
F=-s_j (\gamma P_a +\eta_j {\delta P_a \over \delta s_j}).
\label{flux1}
\end{equation}
Equations (\ref{za}) and  (\ref{pa})  where previously obtained and
studied in \cite{cln} in for the four-wave systems.
The only difference for the four-wave case was different
expressions for $\eta$ and $\gamma$. For the three-wave case,
equation for the PDF  was not considered before, but equations
for its moments were derived and solved in \cite{ln}.
In particular, equation for the first moment is nothing but the
familiar kinetic equation $\dot n = - \gamma n + \eta$ which
gives $\eta = \gamma n$ for any steady state. This, in turn means
that in the steady state with $F=0$ we have
 $P^{(a)}_{j} = (1/n_j) \exp(-s_j /n_j)$ where $n_j$ can be any steady state
solution of th kinetic equation including the KZ spectrum which plays 
the central role in WT \cite{Zakfil,ZLF}.
However, it was shown in \cite{cln} that there also exist 
solutions with $F\ne 0$ which describe WT intermittency.

\section{Phase statistics.}
Importantly, RPA formulation involves independent
{\em phase factors} $\psi = e^\phi$ and not {\em phases}
$\phi$ themselves. Firstly, the phases
would not be convenient because, as we will see later, 
the mean value of the phases is evolving and one could
not say that they are ``distributed uniformly from $-\pi$
to $\pi$''. In fact, we will also see that the mean fluctuation
of the phase distribution is also growing and they quickly 
spread beyond their initial  $2 \pi$-wide interval.
But perhaps even more important, $\phi$'s build mutual correlations 
on the nonlinear 
time whereas $\psi$'s remain independent.
This will be shown later in this section, but
we would like first to give a simple example
illustrating how this property is possible due
to the fact that correspondence between $\phi$
and $\psi$ is not a bijection.

%
%
Let $N$ be a
random integer and let $r_1$ and $r_2$ be
two independent (of $N$ and of each other)
random numbers with uniform distribution between $-\pi$ and
$\pi$. Let
$$\phi_{1,2} = 2 \pi N + r_{1,2}.$$
Then 
$$\langle \phi_{1,2}\rangle=  2 \pi \langle N  \rangle,$$
and
$$ \langle \phi_{1}\phi_{2}\rangle=4\pi^2 \langle N^2 \rangle.$$
Thus,
$$
\langle \phi_{1}\phi_{2}\rangle - \langle \phi_{1} \rangle \langle \phi_{2}\rangle
= 4 \pi^2 ( \langle N^2 \rangle - \langle N \rangle^2) > 0,
$$
which means that variables  $\phi_1$ and
$\phi_2$ are correlated.
On the other hand, if we introduce
$$\psi_{1,2} = e^{i \phi_{1,2}},$$
then
$$\langle \psi_{1,2}\rangle=0,$$
and
$$ \langle \psi_{1}\psi_{2}\rangle=0.$$
$$ \langle \psi_{1}\psi_{2}\rangle-
\langle \psi_{1}\rangle  \langle \psi_{2}\rangle =0,$$
which means that variables  $\psi_1$ and
$\psi_2$ are statistically independent.
In this illustrative example it is clear
that the difference in statistical properties 
between $\phi$ and $\psi$ arises from the fact
that function $\psi(\phi)$ does not have inverse
and, consequently, the information about $N$
contained in $\phi$ is lost in $\psi$.

This illustration, although simple, 
captures the property that actually happens in reality 
as we will show below.
Let us use the following expression for the
phase
$$\phi_j =  \Im \ln {a_j}.$$ Substituting
(\ref{Expansion}) and Taylor-expanding of logarithm in $\epsilon$ one gets

\BEA
\phi_j = \Im\ln(a_j\zer +\e a_j\one +\e^2 a_j\two)
=
\phi\zer +\e \phi\one +\e^2 \phi\two
\EEA
where
\BEA \phi\zer &=& \Im\ln a_j\zer,\\
\phi\one &=&\Im\frac{a_j\one}{a_j\zer},\\ 
\phi\two &=&\Im\left(-\frac{1}{2}\left(\frac{a_j\one}{a_j\zer}\right)^2 +\frac{a_j\two}{a_j\zer}\right). 
\EEA
Now let us perform averaging over the  statistics of  factors $\psi\zer$. 
As usual, the surviving terms are those in which all $\psi\zer$'s cancel out
due to their pairwise matchings. This is possible only if the number of $\psi\zer$'s
is equal to the number of $\bar \psi\zer$'s in the products defining these terms.
Easy to see that the  $\e$ term involves three $\psi\zer$'s
and therefore its average is zero. Therefore,
\BEA \left<\phi_j(T)\right> - \langle\phi\zer_j\rangle 
=
{\e^2 \over A_j^2}
\Im\left<-\frac{1}{2  A_j^2}\left({a_j\one}{\bar a_j\zer}\right)^2 + {a_j\two}{\bar a_j\zer}\right>
\EEA
Let us consider
\BEA
&&\left<(a_j\one\bar a_j\zer)^2\right>  =\left< \sum_{m,n, \kappa, \nu}
\left(V_{mn}^j
a_m a_n\Delta_{mn}^j\delta_{m+n}^j +2\bar V_{jn}^m a_m\bar a_n
\bar\Delta_{jn}^m\delta_{j+n}^m\right)
\left(V_{\kappa \nu}^j
a_\kappa a_\nu\Delta_{\kappa \nu}^j\delta_{\kappa+\nu}^j +2\bar V_{j\nu}^\kappa a_\kappa
\bar a_\nu
\bar\Delta_{jn}^m\delta_{j+n}^m\right)
\bar a_j^2\right> \nonumber\label{phijphijquadrature}
\EEA
Here, there are two terms with equal number of 
$\psi\zer$'s and $\bar \psi\zer$'s but all
couplings of index $j$ to any other index give
zero because $V=0$ if one of its wavenumbers is zero.
Thus, $\left<(a_j\one\bar a_j\zer)^2\right>  =0$.
The other term, $\left<a_j\two\bar a_j\zer\right>_\psi$ has already been
calculated before when evaluating $J_3$. We have
\BEA
\left<a_j\two\bar a_j\zer\right>_\psi =  4\sum_{m,n}
\left[ - |V_{mn}^j|^2\bar E(0,\omega_{mn}^j)\delta_{m+n}^jA_n^2+
|V_{jn}^m|^2 E(0,-\omega_{jn}^m)\delta_{j+n}^m (A_m^2 -A_n^2) 
\right] A_j^2 \nonumber\label{atwoazer}
\EEA
Let us take limits $N \rightarrow \infty$ and $T \rightarrow \infty$ and replace
 $\langle \phi(T)\rangle_\psi -\langle \phi(0)\rangle_\psi /T$ by $\dot {\langle \phi \rangle}_\psi $. 
We get
\BE
\langle \dot\phi_j \rangle_\psi = \omega_{NL},
\EE
where $\omega_{NL}$ is the nonlinear frequency correction
given by
\BE
\omega_{NL}
=
4\e^2\int  \left[|V_{mn}^j|^2 {\cal P}
\left(\frac{1}{\omega_{mn}^j} \right) \delta_{m+n}^j -
|V_{jn}^m|^2 {\cal P} \left(\frac{1}{\omega_{jn}^m} \right)
\delta_{j+n}^m(A_m^2-A_n^2)\right]A_j^2 \, dk_m dk_n
\EE
Here ${\cal P}(x)$ denotes the principal value of the integral. 
Averaging over the amplitudes, we have
$$
\dot {\langle \phi_j \rangle} = \langle \omega_{NL} \rangle,
$$
where $\langle \omega_{NL} \rangle$ is the amplitude-averaged
nonlinear frequency correction
\BE
\langle \omega_{NL} \rangle
=
4\e^2\int  \left[|V_{mn}^j|^2{\cal P}
\left(\frac{1}{\omega_{mn}^j}\right) \delta_{m+n}^j -
|V_{jn}^m|^2 {\cal P} \left(\frac{1}{\omega_{jn}^m} \right) 
\delta_{j+n}^m(n_m-
n_n)\right]n_j \, dk_m dk_n
\EE
We can see that the mean value of the phase is steadily changing
over the nonlinear time and, therefore, it would be incorrect to
assume that the phase ``remains uniformly distributed from $- \pi$
to $\pi$'' even though this could be true for $t=0$.
This is one of the reasons why we formulate RPA in terms of
$\psi$ and not $\phi$. Indeed, $\psi$ was shown above to stay
uniformly distributed on the unit circle over the nonlinear time.

The other reason is that, strictly speaking, $\phi$'s do not
stay de-correlated where as $\psi$'s do (as shown before).
We already saw in the beginning of this section that this
situation is possible due to the fact that the map $\phi \to
\psi = e^{i \phi}$ is not a bijection. Let us now
study such a buildup in statistical dependence of the phases,
let us consider correlator 
${\cal F}_{j,k} \equiv \langle(\phi_j-\langle\phi_j\rangle)
(\phi_k-\langle\phi_k\rangle)\rangle=\langle\phi_j \phi_{k}\rangle-
\langle\phi_j\rangle\langle\ \phi_{k}\rangle.$
At time $T$ we have
\BE
{\cal F}_{j,k}(T)={\cal F}_{j,k}\zer+\e{\cal F}_{j,k}\one+\e^2{\cal F}_{j,k}\two\label{phasecorrelator}
\EE
where
\BEA
{\cal F}_{j,k}\zer &=& \langle\phi_j\zer\phi_k\zer\rangle-
\langle\phi_j\zer\rangle\langle\phi_k\zer\rangle,
\CR
{\cal F}_{j,k}\one &=& \langle\phi_j\one\phi_k\zer\rangle +
\langle\phi_k\one\phi_j\zer\rangle,
\CR
{\cal F}_{j,k}\two &=& \langle\phi_j\one\phi_k\one\rangle +
\langle\phi_j\two\phi_k\zer\rangle +\langle\phi_k\two\phi_j\zer\rangle-
\langle\phi_j\two\rangle\langle\phi_k\zer\rangle-\langle\phi_k\two\rangle\langle\phi_j\zer\rangle.
\EEA
Here, we have taken into account that, as we showed earlier,
$\langle\phi_j\one \rangle =0$.
Let us consider the $\e$-term ${\cal F}_{j,k}\one$, e.g.
\BE
\langle\phi_j\one\phi_k\zer\rangle = \Im \langle (a_j\one\bar a_j\zer)\phi_k\zer \rangle
=
\Im \sum_{k,m,n}V_{mn}^j\langle a_ma_n\bar a_j\phi_k\rangle\Delta_{mn}^j\delta_{m+n}^j 
+2\bar V_{jn}^m\langle a_m\bar a_n\bar a_j\phi_k\rangle\bar 
\Delta_{jn}^m\delta_{j+n}^m\label{jbarjk} 
\EE
In this expression, we have a factor $\phi_k$ which enters directly and not via the
combination $\psi_k = e^{ i \phi_k}$. Potentially, this could greatly complicate the situation
because to objects like $\left< \psi_k \phi_k \right>$ knowledge of the statistics of $\psi$
is not sufficient and one needs the full PDF of $\phi_k$.  
Fortunately, however, this does not cause problems here because, no matter what index
is matched to $k$, matching of the two remaining indices results in $V=0$.
Therefore, the contribution of the $\e$-terms is nill.

Let us now consider the ${\cal F}_{j,k}\two$ starting with
\BEA
\langle\phi_j\two\phi_k\zer\rangle_\psi =
 \left<\Im\left[-\frac{1}{2}\left(\frac{a_j\one}{a_j\zer}\right)^2 + 
\frac{a_j\two}{a_j\zer}\right]\phi_k\zer\right> 
=\Im \left<\left[-\frac{1}{2A_j^4}(a_j\one\bar a_j\zer)^2
+\frac{1}{A_j^2}a_j\two\bar a_j\zer  \right]\phi_k\zer\right>. 
\EEA
We see that the square bracket on the RHS involves an even number (four or six)
of $\psi's$ in each term. Thus, in order for these terms to survive these
$\psi's$ must cancel out which is possible when their indices match in a pairwise
way. But this means that index $k$ (of $\phi_k$) does not match to any
of the indices of $\psi's$ and, therefore, the averaging of  $\phi_k$
can be taken separately because it is statistically independent of 
all other phase factors.\footnote{There is of course also a possibility
that $k$ couples simultaneously to both indices in a pair, but
this contribution contains $\sim 1/N$ less terms and, therefore, 
should be ignored.}
Thus, we conclude that $\langle\phi_j\two\phi_k\zer\rangle_\psi =
\langle\phi_j\two \rangle_\psi \langle \phi_k\zer\rangle_\psi $
and these terms drop out of ${\cal F}_{j,k}\two$.
The remaining term in ${\cal F}_{j,k}\two$ is
\BEA
\langle\phi_j\one\phi_k\one\rangle_\psi &=& 
-\frac{1}{4}\left<\left(\frac{a_j\one}{a_j\zer}-\frac{\bar a_j\one}{\bar a_j\zer}\right)
\left(\frac{a_j\one}{a_j\zer}-\frac{\bar a_j\one}{\bar a_j\zer}\right)\right> 
= -\frac{1}{4}\left<\frac{a_j\one a_k\one}{a_j\zer a_k\zer}
-\frac{a_j\one\bar a_k\one}{a_j\zer a_k\zer} + c.c \right>\CR
&=& 2 |V_{k(j-k)}^j|^2|\Delta_{k(j-k)}^j|^2
A_{j-k}^2
+2 |V_{j(k-j)}^k|^2|\Delta_{j(k-j)}^k|^2
A_{k-j}^2
+ 
{2} |V_{jk}^{j+k}|^2 |\Delta_{jk}^{j+k}|^2 
A_{j+k}^2 \nonumber \\
&+& {\delta^j_k  \over A_{j}^2}
 \sum_{l,m}
\left[|V_{lm}^{j}|^2 |\Delta_{lm}^{j}|^2 \delta_{l+m}^{j} +
2 |V_{jl}^{m}|^2 |\Delta_{jl}^{m}|^2 \delta_{l+j}^{m}
\right]
A_l^2A_m^2
\EEA
We can now average over the amplitudes and
take limits $N \to \infty$ and $\e \to 0$ and write
\BEA
\dot {\cal F}_{j,k} &=& 4 \pi \e^2 \left[
  |V_{k(j-k)}^j|^2 \delta(\omega_{k(j-k)}^j)
n_{j-k}
+  |V_{j(k-j)}^k|^2 \delta(\omega_{j(k-j)}^k)
n_{k-j}
+ 
 |V_{jk}^{j+k}|^2  \delta(\omega_{jk}^{j+k})
n_{j+k} \right]  \nonumber \\
&+& {2 \pi \e^2 \delta^j_k  \over n_{j}}
 \int 
\left[|V_{lm}^{j}|^2 \delta(\omega_{lm}^{j})  \delta_{l+m}^{j} +
2 |V_{jl}^{m}|^2 \delta(\omega_{jl}^{m}) \delta_{l+j}^{m}
\right]
n_l n_m \; dk_l dk_m.
\label{phascor}
\EEA
Presence of the 1-st term on the RHS indicates
 that the phases of the $j$-th and the $k$-th
modes get correlated on the nonlinear time.
This correlation is week in a sense that 
$ {\cal F}_{j,k} $ has a sharp peak at $j=k$ but
the integrated contribution of all $j \ne k$ 
is of the same order as the value at the contribution
of the $j=k$ peak and, therefore, could cause a problem
should one tried to build RPA based on the statistics 
of $\phi$'s rather than $\psi$'s (which remain de-correlated).

Let us consider a special case of
(\ref{phascor}) for $j=k$ which is interesting because it
allows one to calculate the dispersion in phases,
$$\sigma_k = \langle \phi_k^2\rangle - \langle \phi_k\rangle^2$$
We have
\BE
\dot \sigma_k = \eta_k /n_k,
\EE
where $\eta_k$ is defined in (\ref{RHO}) and $n_k=\langle |a_k|^2 \rangle $.
 One can see that the
RHS here is always positive and, therefore, the phase fluctuations 
experience an unlimited growth. On stationary spectra, this 
growth is $\sim \sqrt{t}$ which corresponds to $\sigma \sim t$.
Recall that the mean value of the phase is also changing in time
with the rate $\omega_{NL}$ and on stationary spectra this 
change is linear in time.

\section{ Discussion }

In the present paper, we considered evolution of the full
N-mode objects such as the generating functional and the
probability density function for all the wave amplitudes and 
their phase factors. We proved that the phase factors, being
statistically independent and uniform on $S^1$ initially, remain
so over the nonlinear evolution time in the leading order in
small nonlinearity.
If in addition the initial amplitudes are independent too, then
they remain so over the nonlinear time in a weak sense.
Namely, all joint PDF's for the number of modes $M \ll N$ 
split into products of the one-mode densities
with $O(M/N)$ and $O(\e^2)$ accuracy. Thus, the full $N$-mode PDF
does not factorise as a product of $N$ one-mode densities
and the Fourier modes in the set considered as a whole are not
independent. However, the wave turbulence closure only deals
with the joint objects of the finite size $M$ of variables
while taking $N \to \infty$ limit. These objects do factorise
into products and, for the WT purposes, the Fourier
modes can be interpreted as statistically independent.
In particular, the derivation of the kinetic equation for the
energy spectrum deals only with the $1$-mode and the $2$-mode distributions
and is, therefore, justified by the results of the present paper.
Generally speaking, our results reduce the leading-order WT problem to the
study of the one-mode amplitude PDF's and they validate
 the generalised RPA technique introduced
in \cite{ln,cln}. Such a study of the one-mode PDF and the
high-order momenta of the wave amplitudes was done
in \cite{ln,cln}. It was shown, in particular,
that anomalous probabilities of large wave
amplitudes can appear in the form of a finite-flux solution
in the amplitude space caused by a wave-breaking amplitude cutoff.
The reader is referred to these papers
for the discussion of the WT intermittency.

Although our results indicate that correlations between 2 or more
(but $\ll N$) modes do not appear in the leading (i.e. $\e^2$) order
for the three-wave systems, they definitely appear as corrections
in the next (i.e. $\e^4$) order. Our paper is concerned with the main
order statistics only in which the main evolution happens in the
$1$-mode objects, e.g. the $1$-mode amplitude distributions.
For study of the multi-mode correlations developing in WT in the
next order in $\e$
the reader is referred to papers \cite{NazarenkoNewell,jansen}.

We have also considered correlators of the phase and we
showed the relation between the statistical properties of the
phase $\phi$ and the phase factors $\psi = e^{i\phi}$. 
We showed that  the mean of $\phi$ and its fluctuations
about the mean grow in time and, therefore, there exist
no $2 \pi$-wide interval in which the phase would remain 
uniformly distributed. Moreover, phases $\phi$ become correlated
at different wavenumbers that lie on the resonant manifold.
These properties make the phase $\phi$ an inappropriate
variable for formulating the RPA method of WT description.
On the other hand, our work shows that the   phase factors $\psi = e^{i\phi}$
do remain statistically independent and uniform on $S^1$ which
makes them the right choice for the RPA formulation.

The present paper deals with the three-wave systems only.  The
four-wave resonant interactions are slightly more complicated in
that the nonlinear frequency shift occurs at a lower order in
nonlinearity parameter than the nonlinear evolution of the wave
amplitudes. To build a consistent description of the amplitude
moments one has to perform a renormalisation of the perturbation
series taking into account the nonlinear frequency shift.  This
derivation will be published separately, whereas here we just
announce its main result, the 4-wave generalisation of the
Peierls equation for the PDF. It has the same continuity equation
form (\ref{peierls}) but now the probability flux is
\BEA
F_j &=& 4\pi\e^4\sum_{123}
W_{23}^{j1}\delta(\tilde\omega^{ji}_{23})\delta^{j1}_{23}s_1s_2s_3s_j(\frac{\delta}{\delta
s_j} +\frac{\delta}{\delta s_1} -\frac{\delta}{\delta s_2}
-\frac{\delta}{\delta s_3})P,
\label{gzs}
\EEA
where $W_{23}^{j1}$ is the 4-wave ineraction coefficient and
$\tilde \omega^{l\alpha}_{\mu\nu}
 = \omega^{l\alpha}_{\mu\nu}
 +\Omega_l+\Omega_\alpha-\Omega_\mu-\Omega_\nu$ with  $\Omega_l = 2 \epsilon \sum_\mu
 W^{l \mu}_{l \mu} n_\mu $ being the nonlinear frequency shift.
As wee see this equation is even more compact than its 3-wave analog.
In addition to the derivation of this equation, we will also analyse its
 properties and consequences for the mode correlations and intermittency in
 4-wave  turbulent systems. 


\section{Appendix 1}

Let us obtain $Z(T)$ in terms of the series in small nonlinearity
up to the second order in $\e$.
As an intermediate step, we first consider
separately the amplitude and the phase ingredients of $Z$ and
substitute the $\epsilon$-expansion of $a$ from (\ref{Expansion}) into
their expressions,
\BEA &&e^{\lambda_j |a_j|^2}= e^{\lambda_j |a_j\zer +\e a_j\one
+\e^2 a_j\two|^2}= e^{\lambda_j|a_j\zer|^2 +\e
\lambda_j(a_j\one\bar a_j\zer + \bar a_j\one a_j\zer) +
\e^2\lambda_j\left[|a_j\one|^2+(a_j\two \bar a_j\zer + \bar
a_j\two a_j\zer)\right]}\CR &=& e^{\lambda_j|a_j\zer|^2}\left\{
1+\e \lambda_j(a_j\one\bar a_j\zer + \bar a_j\one a_j\zer) +
\e^2\lambda_j\left[|a_j\one|^2+(a_j\two \bar a_j\zer + \bar
a_j\two a_j\zer)\right]+\frac{\e^2\lambda_j^2}{2}(a_j\one\bar
a_j\zer + \bar a_j\one a_j\zer)^2 \right\}\CR 
&=&
e^{\lambda_j A_j^{(0) 2}} ( 1+\e {\alpha_{1j}} + \e^2
{\alpha_{2j}}),
\label{alpha}\EEA
%
and
\BEA &&\psi_j^{\mu_j}=\left(\frac{a_j\zer +\e a_j\one +\e^2
a_j\two}{\bar a_j\zer +\e \bar a_j\one +\e^2 \bar
a_j\two}\right)^{\frac{\mu_j}{2}}=\psi_j^{(0)\mu_j}\left(\frac{1+\e\frac{a_j\one}{a_j\zer}
+\e^2\frac{a_j\two}{a_j\zer}}{1+\e\frac{\bar a_j\one}{\bar
a_j\zer} +\e^2\frac{\bar a_j\two}{\bar
a_j\zer}}\right)^{\frac{\mu_j}{2}}\CR &=&
\psi_j^{(0)\mu_j} \left[1+\e\frac{\mu_j}{2}\frac{a_j\one}{a_j\zer}
+\e^2\frac{\mu_j}{2}\frac{a_j\two}{a_j\zer}
+\frac{\e^2}{2}\frac{\mu_j}{2}\left(\frac{\mu_j}{2}-1\right)\left(\frac{a_j\one}{a_j\zer}\right)^2\right]
\left[1-\e\frac{\mu_j}{2}\frac{a_j\one}{a_j\zer}
-\e^2\frac{\mu_j}{2}\frac{a_j\two}{a_j\zer}
+\frac{\e^2}{2}\frac{\mu_j}{2}\left(\frac{\mu_j}{2}+1\right)\left(\frac{a_j\one}{a_j\zer}\right)^2\right]\CR
&=& \psi_j^{(0)\mu_j}\left\{ 1+\e
\frac{\mu_j}{2}\left(\frac{a_j\one}{a_j\zer}-\frac{\bar
a_j\one}{\bar a_j\zer}\right) +\e^2\left[\frac{\mu_j}{2}\left(
\frac{a_j\two}{a_j\zer}-\frac{\bar a_j\two}{\bar a_j\zer} \right)
+ \frac{\mu_j}{4}\left[\left(\frac{\mu_j}{2}-1
\right)\left(\frac{a_j\one}{a_j\zer}\right)^2 +
\left(\frac{\mu_j}{2}+1\right)\left(\frac{a_j\one}{a_j\zer}\right)^2\right]-\frac{\mu_j^2|a_j\one|^2}{4A_j^{(0)2}}\right]
\right\}
\CR &=& \psi_j^{(0)\mu_j} (1+\e {\beta_{1j}}
+\e^2 {\beta_{2j}} ),
\label{beta}\EEA
where ${\alpha_{1j}},{\alpha_{2j}},{\beta_{1j}}$ and ${\beta_{2j}}$ denote the
linear and quadratic contributions into the amplitude and phase parts of $Z$ respectively,
\BEA
{\alpha_{1j}} &=& \lambda_j(a_j\one\bar a_j\zer + \bar a_j\one
a_j\zer),  \\
 {\alpha_{2j}} &=&  {(\lambda_j +\lambda_j^2
A_j^{(0)2}|a_j\one|^2 +\lambda_j(a_j\two \bar a_j\zer + \bar a_j\two
a_j\zer)+\frac{\lambda_j^2}{2}(a_j\one \bar a_j\zer)^2 + (\bar
a_j\one a_j\zer)^2}, \\
{\beta_{1j}} &=&
\frac{\mu_j}{2A_j^{(0)2}}(a_j\one\bar a_j\zer-\bar a_j\one
a_j\zer), \\
{\beta_{2j}} &=& \frac{\mu_j}{2A_j^{(0)2}}(a_j\two\bar
a_j\zer-\bar a_j\two a_j\zer)+\frac{\mu_j}{4}\left[
\left(\frac{\mu_j}{2}-1\right)\left(\frac{a_j\one}{a_j\zer}\right)^2
+\left(\frac{\mu_j}{2}+1\right)\left(\frac{\bar a_j\one}{\bar
a_j\zer}\right)^2
\right]-\frac{\mu_j^2|a_j\one|^2}{4A_j^{(0)2}}.
\EEA
Substituting expansions (\ref{alpha}) and (\ref{beta}) into the
expression for $Z$, we have 
%
%

\BEA Z(T)&=&  {1 \over (2 \pi)^{N}}  \left<\prod_{l}
e^{\lambda_l|a_l|^2}\left(\frac{a_l}{\bar
a_l}\right)^{\frac{\mu_l}{2}}\right> =
{1 \over (2 \pi)^{N}} 
 \left<\prod_{l}
e^{\lambda_l A_l^{(0)2}}[1+\e \alpha_{1l}+\e^2
\alpha_{2l}]\psi_l^{(0)\mu_l}[1+\e
\beta_{1l}+\e^2\beta_{2l}]\right>     \CR 
&=&
{1 \over (2 \pi)^{N}} 
\left<\prod_l e^{\lambda_l
A_l^{(0)2}}\psi_l^{(0)\mu_l}\left[1+\e
\sum_j(\alpha_{1j}+\beta_{1j})+\e^2\sum_j(\alpha_{2j}+\beta_{2j})
+\e^2\sum_{j<k}(\alpha_{1j}\alpha_{1k}+\beta_{1j}\beta_{1k})
+\sum_{j,k}\alpha_{1j}\beta_{1k}\right]\right>\CR
&=& 
 {1 \over (2 \pi)^{N}} 
 \left<\prod_l e^{\lambda_l
A_l^{(0)2}}\psi_l^{(0)\mu_l} \big[1+\e\underbrace{\sum_j(\alpha_{1j}+\beta_{1j})}_{I_1}
+\e^2\underbrace{\sum_j(\alpha_{2j}+\beta_{2j}+\alpha_{1j}\beta_{1j})}_{I_2}
+\e^2\underbrace{{1 \over 2} 
\sum_{j \ne k}(\alpha_{1j}\alpha_{1k}+\beta_{1j}\beta_{1k}+2\alpha_{1j}\beta_{1k})}_{I_3}
\big]\right>, \label{Z-expansion} \nonumber \EEA
%
For parts $I_1, I_2$ and
$I_3$ in the above expression we have,
\BEA I_1 &=& \sum_j (\lambda_j
+\frac{\mu_j}{2A_j^2})a_j\one\bar a_j\zer +(\lambda_j
-\frac{\mu_j}{2A_j^2})\bar a_j\one a_j\zer, \nonumber \EEA 
\BEA
I_2 &=& \sum_j(\lambda_j
+\lambda_j^2A_j^2-\frac{\mu_j^2}{2A_j^2})|a_j\one|^2 +
(\lambda_j + \frac{\mu_j}{2A_j^2})a_j\two\bar a_j\zer + (\lambda_j
-\frac{\mu_j}{2A_j^2})\bar a_j\two a_j\zer\CR &+&
\left[\frac{\lambda_j^2}{2}+\frac{\mu_j}{4A_j^4}\left(\frac{\mu_j}{2}-1\right)+\frac{\lambda_j
\mu_j}{2A_j^2} \right](a_j\one \bar a_j\zer)^2 +
\left[\frac{\lambda_j^2}{2}+\frac{\mu_j}{4A_j^4}\left(\frac{\mu_j}{2}+1\right)-\frac{\lambda_j
\mu_j}{2A_j^2} \right](\bar a_j\one a_j\zer)^2, \nonumber \EEA 
\BEA
I_3 &=& {1 \over 2} \sum_{j \ne k} \big[ \lambda_j\lambda_k(a_j\one\bar a_j\zer +\bar
a_j\one a_j\zer)(a_k\one\bar a_k\zer +\bar a_k\one a_k\zer) +
\frac{ \lambda_j \mu_k}{A_k^2}(a_j\one\bar
a_j\zer +\bar a_j\one a_j\zer)(a_k\one\bar a_k\zer -\bar a_k\one
a_k\zer)  \nonumber \\
&+& 
\frac{\mu_j \mu_k}{4A_j^2 A_k^2}(a_j\one\bar
a_j\zer - \bar a_j\one a_j\zer)(a_k\one\bar a_k\zer -\bar a_k\one
a_k\zer) \big], \nonumber 
\EEA

Exploiting the
property
$\bar Z \{\lambda, -\mu\} = Z\{\lambda, \mu\}$ we can write
\BE
 Z\{\lambda, \mu\} =  X\{\lambda, \mu\} +  \bar X \{\lambda, - \mu\}. 
\label{zx1}
\EE
At $t=T$ we have for $ X\{\lambda, \mu\} $
\BE
 X(T) =  X(0) +  (2 \pi)^{2N} \left<\prod_{\|l\|<N}
e^{\lambda_l|a_l\zer|^2}[\e J_1 +\e^2(J_2 +J_3+J_4+J_5)] \right>_A, 
\label{xt1}
\EE
where
\BEA
J_1 &=&   \left<\prod_l \psi_l^{(0)\mu_l} 
\sum_j (\lambda_j
+\frac{\mu_j}{2|a_j\zer|^2})a_j\one\bar a_j\zer \right>_\psi, \label{j11} \\
J_2 &=&   {1 \over 2} \left<\prod_l \psi_l^{(0)\mu_l} 
\sum_j (\lambda_j+
\lambda_j^2|a_j\zer|^2-\frac{\mu_j^2}{2|a_j\zer|^2})|a_j\one|^2
 \right>_\psi, \label{j21} \\
J_3 &=&   \left<\prod_l \psi_l^{(0)\mu_l} 
 \sum_j 
(\lambda_j + \frac{\mu_j}{2|a_j\zer|^2})a_j\two\bar a_j\zer 
 \right>_\psi, \label{j31} \\
J_4 &=&   \left<\prod_l \psi_l^{(0)\mu_l} 
\sum_j 
\left[\frac{\lambda_j^2}{2}+\frac{\mu_j}{4|a_j\zer|^4}\left(\frac{\mu_j}{2}-1\right)+\frac{\lambda_j
\mu_j}{2|a_j\zer|^2} \right](a_j\one \bar a_j\zer)^2
 \right>_\psi, \label{j41} \\
J_5 &=&   {1 \over 2} \left<\prod_l \psi_l^{(0)\mu_l} 
\sum_{j \ne k}\lambda_j\lambda_k(a_j\one\bar a_j\zer +\bar a_j\one
a_j\zer)a_k\one\bar a_k\zer + (\lambda_j
+\frac{\mu_j}{4|a_j\zer|^2})\frac{\mu_k}{|a_k\zer|^2}(a_k\one\bar a_k\zer
-\bar a_k\one a_k\zer)a_j\one\bar a_j\zer
 \right>_\psi, \label{j51} 
\EEA
where $\left< \cdot \right>_A$ and $\left< \cdot \right>_\psi$ denote
the averaging over the initial amplitudes and initial phases
respectively.  We remind that such individual averages are possible
because the amplitudes and the phases are statistically independent
from each other at $t=0$.


\section{Appendix 2}


\subsection{Calculation of $J_3$}


\BEA J_3 &=&  \left<\prod_l \psi_l^{(0)\mu_l} 
 \sum_j 
(\lambda_j + \frac{\mu_j}{2A_j^2})a_j\two\bar a_j\zer 
 \right>_\psi \label{j3p}\CR &=& \langle \prod_l \psi_l^{(0)\mu_l} 
 \sum_j 
(\lambda_j + \frac{\mu_j}{2A_j^2})\CR 
&&\sum_{j,m,n, \kappa, \nu} \left[ 2
V^j_{mn} \left( -V^m_{\kappa \nu}a_n a_\kappa a_\nu E[\omega^j_{n \kappa
\nu},\omega^j_{mn}] \delta^m_{\kappa + \nu} -2 \bar V^\kappa_{m \nu}a_n
a_\kappa \bar a_\nu \bar E[\omega^{j \nu}_{n
\kappa},\omega^j_{mn}]\delta^\kappa_{m + \nu}\right) \delta^j_{m+n}
\right.\CR && \left. + 2 \bar V^m_{jn}
 \left(-V^m_{\kappa \nu}\bar a_n a_\kappa a_\nu E[\omega^{jn}_{\kappa \nu},-\omega^m_{jn}]
\delta^m_{\kappa + \nu} - 2 \bar V^\kappa_{m \nu}\bar a_n a_\kappa \bar
a_\nu E[-\omega^\kappa_{n \nu j},-\omega^m_{j n}]  \delta^\kappa_{m +
\nu} \right) \delta^m_{j+ n} \right. \CR && \left. + 2 \bar
V^m_{jn} \left( \bar V^n_{\kappa \nu}a_m \bar a_\kappa  \bar a_\nu
\delta^n_{\kappa + \nu} E[-\omega^m_{j\nu\kappa},-\omega^m_{jn}] + 2
V^\kappa_{n \nu}a_m \bar a_\kappa  a_\nu E[\omega^{\kappa j}_{\nu m},
-\omega^m_{jn}]\delta^\kappa_{n + \nu}\right)\delta^m_{j+n}
\right] \bar a_j  \rangle_\psi.
 \EEA
The terms to be averaged here can be drawn as
\vskip 1cm

\

\BEA
\parbox{35mm} {
\begin{fmffile}{n30}
   \begin{fmfgraph*}(60,45)
\fmfforce{(0.w,1.h)}{i1}
\fmfforce{(0.w,0.h)}{i2}
\fmfforce{(1.w,1.h)}{o1}
\fmfforce{(1.w,0.h)}{o2}
\fmfforce{(0.25w,0.5h)}{v1}
\fmfforce{(0.75w,0.5h)}{v2}
\fmfforce{(0.5w,1.h)}{v3}
    \fmflabel{$j$}{i1}
    \fmflabel{$n$}{i2}
    \fmflabel{$\kappa$}{o1}
    \fmflabel{$\nu$}{o2}
    \fmf{dashes_arrow}{i1,v1}
    \fmf{dashes_arrow}{v1,i2}
    \fmf{dashes_arrow}{v2,o1}
    \fmf{dashes_arrow}{v2,o2}
\fmf{dots_arrow, label=$m$}{v1,v2}
   \end{fmfgraph*}
\end{fmffile}
}
+ \quad
\parbox{35mm} {
\begin{fmffile}{n31}
   \begin{fmfgraph*}(60,45)
\fmfforce{(0.w,1.h)}{i1}
\fmfforce{(0.w,0.h)}{i2}
\fmfforce{(1.w,1.h)}{o1}
\fmfforce{(1.w,0.h)}{o2}
\fmfforce{(0.25w,0.5h)}{v1}
\fmfforce{(0.75w,0.5h)}{v2}
\fmfforce{(0.5w,1.h)}{v3}
    \fmflabel{$j$}{i1}
    \fmflabel{$n$}{i2}
    \fmflabel{$\kappa$}{o1}
    \fmflabel{$\nu$}{o2}
    \fmf{dashes_arrow}{i1,v1}
    \fmf{dashes_arrow}{v1,i2}
    \fmf{dashes_arrow}{v2,o1}
    \fmf{dashes_arrow}{o2,v2}
\fmf{dots_arrow, label=$m$}{v1,v2}
   \end{fmfgraph*}
\end{fmffile}
}
+ \quad
\parbox{35mm} {
\begin{fmffile}{n32}
   \begin{fmfgraph*}(60,45)
\fmfforce{(0.w,1.h)}{i1}
\fmfforce{(0.w,0.h)}{i2}
\fmfforce{(1.w,1.h)}{o1}
\fmfforce{(1.w,0.h)}{o2}
\fmfforce{(0.25w,0.5h)}{v1}
\fmfforce{(0.75w,0.5h)}{v2}
\fmfforce{(0.5w,1.h)}{v3}
    \fmflabel{$j$}{i1}
    \fmflabel{$n$}{i2}
    \fmflabel{$\kappa$}{o1}
    \fmflabel{$\nu$}{o2}
    \fmf{dashes_arrow}{i1,v1}
    \fmf{dashes_arrow}{i2,v1}
    \fmf{dashes_arrow}{v2,o1}
    \fmf{dashes_arrow}{v2,o2}
\fmf{dots_arrow, label=$m$}{v1,v2}
   \end{fmfgraph*}
\end{fmffile}
}
\nonumber
\EEA
\\
\BEA
+ \quad
\parbox{35mm} {
\begin{fmffile}{n33}
   \begin{fmfgraph*}(60,45)
\fmfforce{(0.w,1.h)}{i1}
\fmfforce{(0.w,0.h)}{i2}
\fmfforce{(1.w,1.h)}{o1}
\fmfforce{(1.w,0.h)}{o2}
\fmfforce{(0.25w,0.5h)}{v1}
\fmfforce{(0.75w,0.5h)}{v2}
\fmfforce{(0.5w,1.h)}{v3}
    \fmflabel{$j$}{i1}
    \fmflabel{$n$}{i2}
    \fmflabel{$\kappa$}{o1}
    \fmflabel{$\nu$}{o2}
    \fmf{dashes_arrow}{i1,v1}
    \fmf{dashes_arrow}{i2,v1}
    \fmf{dashes_arrow}{v2,o1}
    \fmf{dashes_arrow}{o2,v2}
\fmf{dots_arrow, label=$m$}{v1,v2}
   \end{fmfgraph*}
\end{fmffile}
}
+ \quad
\parbox{35mm} {
\begin{fmffile}{n34}
   \begin{fmfgraph*}(60,45)
\fmfforce{(0.w,1.h)}{i1}
\fmfforce{(0.w,0.h)}{i2}
\fmfforce{(1.w,1.h)}{o1}
\fmfforce{(1.w,0.h)}{o2}
\fmfforce{(0.25w,0.5h)}{v1}
\fmfforce{(0.75w,0.5h)}{v2}
\fmfforce{(0.5w,1.h)}{v3}
    \fmflabel{$j$}{i1}
    \fmflabel{$m$}{i2}
    \fmflabel{$\kappa$}{o1}
    \fmflabel{$\nu$}{o2}
    \fmf{dashes_arrow}{i1,v1}
    \fmf{dashes_arrow}{v1,i2}
    \fmf{dashes_arrow}{o1,v2}
    \fmf{dashes_arrow}{o2,v2}
\fmf{dots_arrow, label=$n$}{v2,v1}
   \end{fmfgraph*}
\end{fmffile}
}
+ \quad
\parbox{35mm} {
\begin{fmffile}{n35}
   \begin{fmfgraph*}(60,45)
\fmfforce{(0.w,1.h)}{i1}
\fmfforce{(0.w,0.h)}{i2}
\fmfforce{(1.w,1.h)}{o1}
\fmfforce{(1.w,0.h)}{o2}
\fmfforce{(0.25w,0.5h)}{v1}
\fmfforce{(0.75w,0.5h)}{v2}
\fmfforce{(0.5w,1.h)}{v3}
    \fmflabel{$j$}{i1}
    \fmflabel{$m$}{i2}
    \fmflabel{$\kappa$}{o1}
    \fmflabel{$\nu$}{o2}
    \fmf{dashes_arrow}{i1,v1}
    \fmf{dashes_arrow}{v1,i2}
    \fmf{dashes_arrow}{o1,v2}
    \fmf{dashes_arrow}{v2,o2}
\fmf{dots_arrow, label=$n$}{v2,v1}
   \end{fmfgraph*}
\end{fmffile}
}
\EEA
\vskip 1cm
Let us now average over the random phases.  Again, the leading order
terms will be given by the diagrams with the largest number of
internal couplings.  They will arise from the $V\bar V$ terms (the
2nd, 3rd and the 6th graphs) because they allow 2 internal couplings
in each of them.  There are also possibilities to have one internal
and two external couplings of the dashed lines, - such terms will give
a $O(1/N)$ correction to the leading order.  Therefore,
\BEA
J_3 &=&\left(
\parbox{35mm} {
\begin{fmffile}{n28}
   \begin{fmfgraph*}(70,50) \fmfkeep{theta2}
\fmfforce{(0.1w,0.5h)}{v1}
\fmfforce{(0.9w,0.5h)}{v2}
\fmf{dots_arrow, label=$m$}{v1,v2}
    \fmf{dashes_arrow, right=.7, label= $n$}{v1,v2}
    \fmf{dashes_arrow, right=.7, label= $j$}{v2,v1}
   \end{fmfgraph*}
\end{fmffile}
} 
+2 \hspace{.5cm} 
\parbox{35mm} {
\begin{fmffile}{n29}
   \begin{fmfgraph*}(70,50) \fmfkeep{theta3}
\fmfforce{(0.1w,0.5h)}{v1}
\fmfforce{(0.9w,0.5h)}{v2}
\fmf{dots_arrow, label=$m$}{v1,v2}
    \fmf{dashes_arrow, left=.7, label= $n$}{v2,v1}
    \fmf{dashes_arrow, right=.7, label= $j$}{v2,v1}
   \end{fmfgraph*}
\end{fmffile}
} 
+ \hspace{.5cm} 
\parbox{35mm} {
\begin{fmffile}{n29p}
   \begin{fmfgraph*}(70,50) \fmfkeep{theta4}
\fmfforce{(0.1w,0.5h)}{v1}
\fmfforce{(0.9w,0.5h)}{v2}
\fmf{dots_arrow, label=$n$}{v2,v1}
    \fmf{dashes_arrow, right=.7, label= $m$}{v1,v2}
    \fmf{dashes_arrow, right=.7, label= $j$}{v2,v1}
   \end{fmfgraph*}
\end{fmffile}
} 
\right)[1+O(1/N)]
\nonumber
\EEA
\\
\BEA
&=& 4 \prod_l\delta(\mu_l)
\sum_{j,m,n}
\lambda_j
\left[ - |V_{mn}^j|^2\bar E(0,\omega_{mn}^j)\delta_{m+n}^jA_n^2+
|V_{jn}^m|^2 E(0,-\omega_{jn}^m)\delta_{j+n}^m (A_m^2 -A_n^2) 
\right] A_j^2 
\nonumber \\
&&\hspace{7cm} \times [1+O(1/N)]
\EEA


\subsection{Calculation of $J_4$}


\BEA
J_4 &=&   \left<\prod_l \psi_l^{(0)\mu_l} 
\sum_j 
\left[\frac{\lambda_j^2}{2}+\frac{\mu_j}{4A_j^4}\left(\frac{\mu_j}{2}-1\right)+\frac{\lambda_j
\mu_j}{2A_j^2} \right]
(a_j\one \bar a_j\zer)^2
 \right>_\psi \nonumber \\
&=&\langle \prod_l \psi_l^{\mu_l}
\sum_{j,m,n, \kappa, \nu}
\left[\frac{\lambda_j^2}{2}+\frac{\mu_j}{4A_j^4}\left(\frac{\mu_j}{2}-1\right)+\frac{\lambda_j
\mu_j}{2A_j^2} \right] \nonumber \\
 &&
\left(V_{mn}^j
a_m a_n\Delta_{mn}^j\delta_{m+n}^j +2\bar V_{jn}^m a_m\bar a_n
\bar\Delta_{jn}^m\delta_{j+n}^m\right)
\left(V_{\kappa \nu}^j
a_\kappa a_\nu\Delta_{\kappa \nu}^j\delta_{\kappa+\nu}^j +2\bar V_{j\nu}^\kappa a_\kappa
\bar a_\nu
\bar\Delta_{jn}^m\delta_{j+n}^m\right)
\bar a_j^2\rangle_\psi.
\EEA
Graphically, the 4 terms to be averaged in this expression are
\\
\vskip 1cm
\BEA
\parbox{40mm} {
\begin{fmffile}{n20}
   \begin{fmfgraph*}(80,40)
\fmfforce{(0.w,1.h)}{i1}
\fmfforce{(0.w,0.h)}{i2}
\fmfforce{(1.w,1.h)}{o1}
\fmfforce{(1.w,0.h)}{o2}
\fmfforce{(0.25w,0.5h)}{v1}
\fmfforce{(0.75w,0.5h)}{v2}
\fmfforce{(0.5w,1.h)}{v3}
    \fmflabel{$m$}{i1}
    \fmflabel{$n$}{i2}
    \fmflabel{$\kappa$}{o1}
    \fmflabel{$\nu$}{o2}
    \fmflabel{$j$}{v3}
    \fmf{dashes_arrow}{v1,i1}
    \fmf{dashes_arrow}{v1,i2}
    \fmf{dashes_arrow}{v2,o1}
    \fmf{dashes_arrow}{v2,o2}
\fmf{dashes_arrow}{v3,v1}
\fmf{dashes_arrow}{v3,v2}
   \end{fmfgraph*}
\end{fmffile}
} 
+ \quad
\parbox{40mm} {
\begin{fmffile}{n21}
   \begin{fmfgraph*}(80,40)
\fmfforce{(0.w,1.h)}{i1}
\fmfforce{(0.w,0.h)}{i2}
\fmfforce{(1.w,1.h)}{o1}
\fmfforce{(1.w,0.h)}{o2}
\fmfforce{(0.25w,0.5h)}{v1}
\fmfforce{(0.75w,0.5h)}{v2}
\fmfforce{(0.5w,1.h)}{v3}
    \fmflabel{$m$}{i1}
    \fmflabel{$n$}{i2}
    \fmflabel{$\kappa$}{o1}
    \fmflabel{$\nu$}{o2}
    \fmflabel{$j$}{v3}
    \fmf{dashes_arrow}{v1,i1}
    \fmf{dashes_arrow}{v1,i2}
    \fmf{dashes_arrow}{v2,o1}
    \fmf{dashes_arrow}{o2,v2}
\fmf{dashes_arrow}{v3,v1}
\fmf{dashes_arrow}{v3,v2}
   \end{fmfgraph*}
\end{fmffile}
} 
\nonumber\\ \nonumber\\ \nonumber\\ 
+ \quad
\parbox{40mm} {
\begin{fmffile}{n22}
   \begin{fmfgraph*}(80,40)
\fmfforce{(0.w,1.h)}{i1}
\fmfforce{(0.w,0.h)}{i2}
\fmfforce{(1.w,1.h)}{o1}
\fmfforce{(1.w,0.h)}{o2}
\fmfforce{(0.25w,0.5h)}{v1}
\fmfforce{(0.75w,0.5h)}{v2}
\fmfforce{(0.5w,1.h)}{v3}
    \fmflabel{$m$}{i1}
    \fmflabel{$n$}{i2}
    \fmflabel{$\kappa$}{o1}
    \fmflabel{$\nu$}{o2}
    \fmflabel{$j$}{v3}
    \fmf{dashes_arrow}{v1,i1}
    \fmf{dashes_arrow}{i2,v1}
    \fmf{dashes_arrow}{v2,o1}
    \fmf{dashes_arrow}{v2,o2}
\fmf{dashes_arrow}{v3,v1}
\fmf{dashes_arrow}{v3,v2}
   \end{fmfgraph*}
\end{fmffile}
} 
+ \quad
\parbox{40mm} {
\begin{fmffile}{n23}
   \begin{fmfgraph*}(80,40)
\fmfforce{(0.w,1.h)}{i1}
\fmfforce{(0.w,0.h)}{i2}
\fmfforce{(1.w,1.h)}{o1}
\fmfforce{(1.w,0.h)}{o2}
\fmfforce{(0.25w,0.5h)}{v1}
\fmfforce{(0.75w,0.5h)}{v2}
\fmfforce{(0.5w,1.h)}{v3}
    \fmflabel{$m$}{i1}
    \fmflabel{$n$}{i2}
    \fmflabel{$\kappa$}{o1}
    \fmflabel{$\nu$}{o2}
    \fmflabel{$j$}{v3}
    \fmf{dashes_arrow}{v1,i1}
    \fmf{dashes_arrow}{i2,v1}
    \fmf{dashes_arrow}{v2,o1}
    \fmf{dashes_arrow}{o2,v2}
\fmf{dashes_arrow}{v3,v1}
\fmf{dashes_arrow}{v3,v2}
   \end{fmfgraph*}
\end{fmffile}
} \nonumber \EEA 
\vskip 1cm
Note that there is no dotted lines in these graphs
because for each summation index there is a corresponding wave
amplitude present.  As a consequence, the rule for the number of
surviving summations is somewhat different from what we had so
far. Namely, the number of the summation indices after the phase
averaging is one less than the number of the purely internal
couplings.  Easy to see that the phase averaging of the above terms
always leads to an external coupling of the dashed lines $j$ which
removes the $j$ summation. Moreover, no more than one purely internal
coupling of the dashed lines is possible in any of these
graphs.~\footnote{The only possibility of the double internal coupling
would be in the last graph via joining $m$ with $\nu$ and $\kappa$
with $n$, but this would mean $j=0$ because of the $\delta$-symbols
and, therefore, this term is nill.}  Thus, $J_4$ contains no summation
at all and is only a $O(1/N^2)$ correction to the main terms in $J_2$
and $J_3$.

\subsection{Calculation of $J_5$}
%
Expression for $J_5$ seemingly involves a great number terms.
However, this number can be dramatically reduced by the following
speculation. Just as in $J_4$ there is no dotted lines in the graphs
involved in $J_5$ because for each summation index there is a
corresponding wave amplitude present.  Thus, we have the same rule for
the number of summations surviving the phase averaging (i.e. one less
than the number of internal couplings). In order to be of the same
order as the leading terms in $J_2$ and $J_3$, we must have 3 purely
internal couplings and, therefore, no external couplings. This is only
possible when the number of dotted lines directed to the vertices is
equal to the number of them pointing away which is true for the $V
\bar V$ terms but not true for the $V V$ and $\bar V \bar V$
terms. Thus we will only consider the $V \bar V$ terms. Further, the
fact that there is no external couplings means that such terms are
only non-zero when all $\mu$'s are zero.  Thus, there will be no
contribution from the second part of $J_5$ which has a $\mu_k$
pre-factor.
\BEA
J_5 &=&   {1 \over 2} \left<\prod_l \psi_l^{(0)\mu_l} 
\sum_{j \ne k}\lambda_j\lambda_k(a_j\one\bar a_j\zer +\bar a_j\one
a_j\zer)a_k\one\bar a_k\zer 
 \right>_\psi [1+O(1/N)],  
\EEA
where the  $V \bar V$ terms to be averaged here are
\vskip 1cm
\BEA
C_1 &=& \hspace{1cm} 
\parbox{50mm} {
\begin{fmffile}{n24}
   \begin{fmfgraph*}(100,40)
\fmfforce{(0.w,1.h)}{i1}
\fmfforce{(0.w,0.h)}{i2}
\fmfforce{(1.w,1.h)}{o1}
\fmfforce{(1.w,0.h)}{o2}
\fmfforce{(0.25w,0.5h)}{v1}
\fmfforce{(0.75w,0.5h)}{v2}
\fmfforce{(0.5w,1.h)}{v3}
\fmfforce{(0.5w,0.h)}{v4}
    \fmflabel{$m$}{i1}
    \fmflabel{$n$}{i2}
    \fmflabel{$\kappa$}{o1}
    \fmflabel{$\nu$}{o2}
    \fmflabel{$j$}{v3}
    \fmflabel{$k$}{v4}
    \fmf{dashes_arrow}{v1,i1}
    \fmf{dashes_arrow}{v1,i2}
\fmf{dashes_arrow}{v3,v1}
    \fmf{dashes_arrow}{v2,o1}
    \fmf{dashes_arrow}{o2,v2}
\fmf{dashes_arrow}{v4,v2}
   \end{fmfgraph*}
\end{fmffile}
}
\nonumber
\EEA
\BEA
&& \hspace{5cm} 
= - \prod_l\psi_l^{(0)\mu_l}\sum_{j\neq k,m,n,\mu,\nu}\lambda_j\lambda_k
V_{mn}^j\delta_{m+n}^j a_m a_n \bar a_j \bar V_{k\nu}^\kappa 
\delta_{k+\nu}^\kappa
a_\kappa \bar a_\nu 
\bar a_k \Delta_{mn}^j \bar\Delta_{k\nu}^\kappa , \nonumber 
\EEA
\\
\BEA
C_2 &=& \hspace{1cm} 
\parbox{50mm} {
\begin{fmffile}{n25}
   \begin{fmfgraph*}(100,40)
\fmfforce{(0.w,1.h)}{i1}
\fmfforce{(0.w,0.h)}{i2}
\fmfforce{(1.w,1.h)}{o1}
\fmfforce{(1.w,0.h)}{o2}
\fmfforce{(0.25w,0.5h)}{v1}
\fmfforce{(0.75w,0.5h)}{v2}
\fmfforce{(0.5w,1.h)}{v3}
\fmfforce{(0.5w,0.h)}{v4}
    \fmflabel{$m$}{i1}
    \fmflabel{$n$}{i2}
    \fmflabel{$\kappa$}{o1}
    \fmflabel{$\nu$}{o2}
    \fmflabel{$j$}{v3}
    \fmflabel{$k$}{v4}
    \fmf{dashes_arrow}{v1,i1}
    \fmf{dashes_arrow}{i2,v1}
\fmf{dashes_arrow}{v3,v1}
    \fmf{dashes_arrow}{v2,o1}
    \fmf{dashes_arrow}{v2,o2}
\fmf{dashes_arrow}{v4,v2}
   \end{fmfgraph*}
\end{fmffile}
}
\nonumber
\EEA
\BEA
&& \hspace{5cm} 
= - \prod_l\psi_l^{(0)\mu_l}\sum_{j\neq k,m,n,\mu,\nu}\lambda_j\lambda_k
\bar V_{jn}^m\delta_{j+n}^m a_m \bar a_n \bar a_j 
V_{\kappa \nu}^k \delta_{\kappa+\nu}^k
a_\kappa a_\nu \bar a_k
\bar \Delta_{jn}^m \Delta_{\kappa \nu}^k, \nonumber 
\EEA
\\
\BEA
C_3 &=& \hspace{1cm}
\parbox{50mm} {
\begin{fmffile}{n26}
   \begin{fmfgraph*}(100,40)
\fmfforce{(0.w,1.h)}{i1}
\fmfforce{(0.w,0.h)}{i2}
\fmfforce{(1.w,1.h)}{o1}
\fmfforce{(1.w,0.h)}{o2}
\fmfforce{(0.25w,0.5h)}{v1}
\fmfforce{(0.75w,0.5h)}{v2}
\fmfforce{(0.5w,1.h)}{v3}
\fmfforce{(0.5w,0.h)}{v4}
    \fmflabel{$m$}{i1}
    \fmflabel{$n$}{i2}
    \fmflabel{$\kappa$}{o1}
    \fmflabel{$\nu$}{o2}
    \fmflabel{$j$}{v3}
    \fmflabel{$k$}{v4}
    \fmf{dashes_arrow}{i1,v1}
    \fmf{dashes_arrow}{v1,i2}
\fmf{dashes_arrow}{v1,v3}
    \fmf{dashes_arrow}{v2,o1}
    \fmf{dashes_arrow}{o2,v2}
\fmf{dashes_arrow}{v4,v2}
   \end{fmfgraph*}
\end{fmffile}
}
\nonumber
\EEA
\BEA
&& \hspace{5cm} =
{2}\prod_l\psi_l^{(0)\mu_l}\sum_{j\neq k,m,n,\mu,\nu}\lambda_j\lambda_k
V_{jn}^m \delta_{j+n}^m \bar a_m a_n  a_j \bar V_{k\nu}^\kappa
\delta_{k+\nu}^\kappa
a_\kappa \bar a_\nu \bar a_k \Delta_{jn}^m \bar \Delta_{k\nu}^\kappa,
 \nonumber 
\EEA
\\
\BEA
C_4 &=& \hspace{1cm} 
\parbox{50mm} {
\begin{fmffile}{n27}
   \begin{fmfgraph*}(100,40)
\fmfforce{(0.w,1.h)}{i1}
\fmfforce{(0.w,0.h)}{i2}
\fmfforce{(1.w,1.h)}{o1}
\fmfforce{(1.w,0.h)}{o2}
\fmfforce{(0.25w,0.5h)}{v1}
\fmfforce{(0.75w,0.5h)}{v2}
\fmfforce{(0.5w,1.h)}{v3}
\fmfforce{(0.5w,0.h)}{v4}
    \fmflabel{$m$}{i1}
    \fmflabel{$n$}{i2}
    \fmflabel{$\kappa$}{o1}
    \fmflabel{$\nu$}{o2}
    \fmflabel{$j$}{v3}
    \fmflabel{$k$}{v4}
    \fmf{dashes_arrow}{i1,v1}
    \fmf{dashes_arrow}{i2,v1}
\fmf{dashes_arrow}{v1,v3}
    \fmf{dashes_arrow}{v2,o1}
    \fmf{dashes_arrow}{v2,o2}
\fmf{dashes_arrow}{v4,v2}
   \end{fmfgraph*}
\end{fmffile}
}
\nonumber
\EEA
\BEA
&& \hspace{5cm} =
\frac{1}{2}\prod_l\psi_l^{(0)\mu_l}\sum_{j\neq k,m,n,\mu,\nu}\lambda_j\lambda_k
\bar V_{mn}^j\delta_{m+n}^j \bar a_m \bar a_n a_j V_{\kappa \nu}^k
\delta_{\kappa+\nu}^k
a_\kappa a_\nu\bar a_k\bar \Delta_{mn}^j \Delta_{\kappa \nu}^k. \nonumber 
\EEA

\vskip 1cm
By coupling the dashed lines we have in the leading order

\

\BEA
\langle C_1 \rangle_\psi = &
2  \hspace{.1cm} 
\parbox{35mm} {
\begin{fmffile}{n40}
   \begin{fmfgraph*}(70,50) \fmfkeep{theta5}
\fmfforce{(0.1w,0.5h)}{v1}
\fmfforce{(0.9w,0.5h)}{v2}
\fmf{dashes_arrow, label=$j$}{v2,v1}
    \fmf{dashes_arrow, right=.7, label= $n$}{v1,v2}
    \fmf{dashes_arrow, left=.7, label= $k$}{v1,v2}
   \end{fmfgraph*}
\end{fmffile}
} 
&
= - 2\prod_l\delta(\mu_l) \sum_{j\neq k,n}\lambda_j\lambda_k
|V_{kn}^j|^2 \delta_{k+n}^j A_j^2 A_n^2 A_k^2 
|\Delta_{kn}^j|^2  , \nonumber 
\EEA
\\
\BEA
\langle C_2 \rangle_\psi = &
2  \hspace{.1cm} 
\parbox{35mm} {
\begin{fmffile}{n41}
   \begin{fmfgraph*}(70,50) \fmfkeep{theta6}
\fmfforce{(0.1w,0.5h)}{v1}
\fmfforce{(0.9w,0.5h)}{v2}
\fmf{dashes_arrow, label=$j$}{v2,v1}
    \fmf{dashes_arrow, left=.7, label= $n$}{v2,v1}
    \fmf{dashes_arrow, left=.7, label= $k$}{v1,v2}
   \end{fmfgraph*}
\end{fmffile}
} 
&
= - 2\prod_l\delta(\mu_l) \sum_{j\neq k,n}\lambda_j\lambda_k
|V_{jn}^k|^2 \delta_{j+n}^k A_j^2 A_n^2 A_k^2 
|\Delta_{jn}^k|^2  = \langle C_1 \rangle_\psi, \nonumber 
\EEA
\\
\BEA
\langle C_3 \rangle_\psi =  &
 \hspace{.1cm} 
\parbox{35mm} {
\begin{fmffile}{n42}
   \begin{fmfgraph*}(70,50) \fmfkeep{theta7}
\fmfforce{(0.1w,0.5h)}{v1}
\fmfforce{(0.9w,0.5h)}{v2}
\fmf{dashes_arrow, label=$j$}{v1,v2}
    \fmf{dashes_arrow, right=.7, label= $k$}{v1,v2}
    \fmf{dashes_arrow, right=.7, label= $m$}{v2,v1}
   \end{fmfgraph*}
\end{fmffile}
} 
&
= 2\prod_l\delta(\mu_l) \sum_{j\neq k,m}\lambda_j\lambda_k
|V_{jk}^m|^2 \delta_{j+k}^m A_j^2 A_m^2 A_k^2 
|\Delta_{jk}^m|^2  . \nonumber 
\EEA

\

Term $C_4$ does not survive the averaging because $j \ne k$ and a
triple internal coupling is not possible.  Summarising,
\BE
J_5= 2 \prod_l\delta(\mu_l) \sum_{j\neq k,n}\lambda_j\lambda_k
\left[ 
-2 |V_{kn}^j|^2 \delta_{k+n}^j |\Delta_{kn}^j|^2 
+|V_{jk}^n|^2 \delta_{j+k}^n |\Delta_{jk}^n|^2 
\right]
A_j^2 A_n^2 A_k^2 \;\; [1+O(1/N)].
\EE


\end{document}